%% file: main_v2.tex
\documentclass[journal]{./TeX/IEEEtran}
\input{./TeX/commands.tex}

\begin{document}

%
% paper title
% Titles are generally capitalized except for words such as a, an, and, as,
% at, but, by, for, in, nor, of, on, or, the, to and up, which are usually
% not capitalized unless they are the first or last word of the title.
% Linebreaks \\ can be used within to get better formatting as desired.
% Do not put math or special symbols in the title.
\title{What Does a One-Bit Quanta Image Sensor Offer?}
%\title{Theoretical Analysis of One-bit Quanta Image Sensors: Low-light, Frame Rate, Dynamic Range}
%
%
% author names and IEEE memberships
% note positions of commas and nonbreaking spaces ( ~ ) LaTeX will not break
% a structure at a ~ so this keeps an author's name from being broken across
% two lines.
% use \thanks{} to gain access to the first footnote area
% a separate \thanks must be used for each paragraph as LaTeX2e's \thanks
% was not built to handle multiple paragraphs

\author{Stanley~H.~Chan,~\IEEEmembership{Senior~Member,~IEEE}% <-this % stops a space
\thanks{The author is with the School of Electrical and Computer
Engineering, Purdue University, West Lafayette, IN 47907, USA. Email: {stanchan}@purdue.edu. The work is supported, in part, by the National Science Foundation under the grants IIS-2133032, ECCS-2030570, a gift from Intel Lab, and a gift from Google.}
% <-this % stops a space
%\thanks{}% <-this % stops a space
%\thanks{Manuscript received April 19, 2005; revised August 26, 2015.}
}

\maketitle

% As a general rule, do not put math, special symbols or citations
% in the abstract or keywords.
\begin{abstract}
The one-bit quanta image sensor (QIS) is a photon-counting device that captures image intensities using binary bits. Assuming that the analog voltage generated at the floating diffusion of the photodiode follows a Poisson-Gaussian distribution, the sensor produces either a ``1'' if the voltage is above a certain threshold or ``0'' if it is below the threshold. The concept of this binary sensor has been proposed for more than a decade and physical devices have been built to realize the concept. However, what benefits does a one-bit QIS offer compared to a conventional multi-bit CMOS image sensor? Besides the known empirical results, are there theoretical proofs to support these findings?

The goal of this paper is to provide new theoretical support from a signal processing perspective. In particular, it is theoretically found that the sensor can offer three benefits: (1) Low-light: One-bit QIS performs better at low-light because it has a low read noise and its one-bit quantization can produce an error-free measurement. However, this requires the exposure time to be appropriately configured. (2) Frame rate: One-bit sensors can operate at a much higher speed because a response is generated as soon as a photon is detected. However, in the presence of read noise, there exists an optimal frame rate beyond which the performance will degrade. A Closed-form expression of the optimal frame rate is derived. (3) Dynamic range: One-bit QIS offers a higher dynamic range. The benefit is brought by two complementary characteristics of the sensor: nonlinearity and exposure bracketing. The decoupling of the two factors is theoretically proved, and closed-form expressions are derived.
\end{abstract}

% Note that keywords are not normally used for peerreview papers.

\section{Introduction}
\IEEEPARstart{T}{he} Kodak No. 1 camera produced by George Eastman in 1888 is an important milestone in the human history. With the slogan ``You press the button and we do the rest'', Eastman successfully brought photography to the general public. Certainly, from a technological perspective, the history of photographic films can be traced back to the pioneer work of people such as Henry Talbot (1800-1877) and Louis Daguerre (1787-1851). Historians also give credits to the work of Ferdinand Hurter and Vero Charles Driffield \cite{Clark_1961, Mees_1931} who published an important paper in the \emph{Journal of the Society of Chemical Industry} in 1890 \cite{Hurter_Driffield_1890}. The Hurter-Driffield paper provided the first and the most detailed analysis of how the opacity of the chemicals in a photographic plate would respond to the exposure. According to Hurter and Driffield, the opacity and transparency of a plate are determined by the \emph{density} of the molecules undergoing the photochemical process, and this density is a function of the underlying scene exposure. Because the bonding of the chemicals is either maintained or broken by the incoming light, the measurements recorded by the photographic plate are binary. Using today's terminology in digital signal processing, they are \emph{one-bit}.

\begin{figure}[h]
\centering
\vspace{-2ex}
\includegraphics[width=\linewidth]{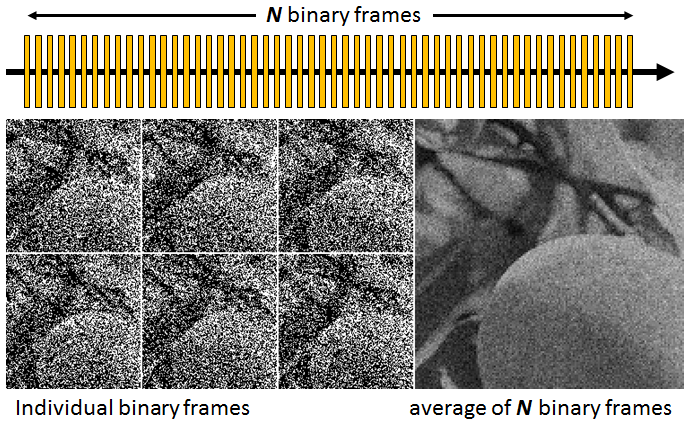}
\vspace{-4ex}
\caption{\textbf{Image formation of a one-bit QIS}. A one-bit quanta image sensor (QIS) captures the scene using many short exposure frames with binary quantization. The example shown here uses $N = 64$ binary frames, simulated from a one-bit Poisson distribution.}
\label{fig: liberty figure}
\vspace{-2ex}
\end{figure}

Fast forward the history to 2005 when CCD and CMOS image sensors became the two dominant technologies in the camera market, Junichi Nakamura and colleagues put together a book \emph{Image Sensors and Signal Processing for Digital Still Cameras} \cite{Nakamura_2005_book}. In the last chapter of the book, Eric Fossum brought a proposal which he considered to be a possible next-generation digital image sensor. This type of sensors, now broadly known as the quanta image sensors (QIS) \cite{Fossum_Proposal_2005, Fossum_SomeThoughts_2006, Fossum_Concept_2011}, uses binary bits to record the scene instead of an analog signal or a 16-bit digital number. Although each pixel is binary, a burst of the one-bit frames contains information to reconstruct the image, as shown in \fref{fig: liberty figure}. In his 2005 paper \cite{Fossum_Proposal_2005}, Fossum called the sensor as a \emph{digital film} because the binary sensing mechanism is reminiscent of the silver halide films. In fact, if one plots the bit density as a function of the exposure (the D-logH curve), one-bit QIS will have a good match with the famous curve generated by Hurter and Driffield back in 1890, as mentioned in \cite[Figure 3]{Fossum_MDPI_2016}.

\subsection{Landscape of Quanta Image Sensors Today}
To achieve one-bit photon counting, the critical technological bottleneck is the deep sub-electron input-referred read noise which determines how well we can count photons at room temperature \cite{Vogelsang_Stork_2012,Vogelsang_Guidash_Xue_2013,Burri_Architecture_2014,Ma_Jot03_2015,Ma_Character_2015,Dutton_HDR_2016,Dutton_Noise_2016,Burri_Monolithic_2018}. Two technologies are currently pursued: using single-photon avalanche diodes (SPAD) by applying a high electric field \cite{Burri_Architecture_2014, Dutton_QVGA_2015, Dutton_Noise_2016}, or by using a high conversion gain transfer device such as a CMOS image sensor (CIS) \cite{Ma_Jot03_2015, Ma_Pump_2015, Ma_Character_2015, Ma_PumpGate_2017,Ma_019e_2021}. The former type of QIS is known as the SPAD-QIS whereas the latter is known the CIS-QIS. SPAD-QIS and CIS-QIS have their respective pros and cons. For example, SPAD-QIS has a zero read noise but a higher dark current. Its speed is superb, but the fill factor, quantum efficiencies and power consumptions are still behind CIS-QIS. CIS-QIS, on the other hand, are making steps towards the consumer electronics market because of its compatibility with the mainstream CMOS fabrication processes. They are particularly appealing for low-light applications for small devices. A comprehensive discussion of the latest development of the sensors can be found in \cite{Ma_review_2022}.

From an image processing point of view, the idea of rapidly capturing short exposure images within a fixed exposure time budget has been explored in the context of CMOS image sensors \cite{Hasinoff_Time_2009, Hasinoff_HDR_2010, Hasinoff_Burst_2016}. Generalization to other multiplexing schemes in the presence of read noise is discussed in \cite{Schechner_Multiplex_2003}. However, as we push the sensing limit to a single bit, the statistics and the theoretical properties of the signals are less obvious. These will be discussed later in the paper.

\subsection{What Theoretical Results are Available?}
Majority of the signal and image processing research for QIS are about solving the inverse problem, i.e., image reconstruction from the binary bits \cite{Chan_PnP_2016,Chan_MDPI_2016,Gyongy_MDPI_2018,Choi_ICASSP_2018,Gnanasambandam_Megapixel_2019,Chi_ECCV_2020, Elgendy_Demosaic_2021,Chengxi_ICCVW_2021}. On the theoretical analysis side, a large portion of the known results are developed for CIS-QIS. They can be summarized in four groups:
\begin{itemize}
\item The statistics of the one-bit signal can be modeled by a truncated Poisson-Gaussian distribution \cite{Fossum_Modeling_2013}. The model can be extended to multi-bit \cite{Fossum_Multi_2015}. The bit density shows insensitivity for appropriately chosen exposure and threshold \cite{Chan_Density_2022}. Equations of the bit error rate have been derived \cite{Fossum_ErrorRate_2016}.
\item In the absence of read noise, the maximum-likelihood estimate using the one-bit signals is asymptotically optimal and can achieve the Cramer-Rao lower bound \cite{Yang_SPIE_2010, Yang_GigaVision_2009, Yang_TIP_2011, Yang_Thesis_2012}.
\item Again, in the absence of read noise, the optimal threshold for an arbitrary exposure is the one that makes the bit-density equal to 0.5 over the space-time volume \cite{Elgendy_Threshold_2018}.
\item Dynamic range of the one-bit sensor is generally wider than a convention CMOS pixel because one-bit sensors can perform multiple exposures bracketing \cite{Yang_Thesis_2012,Lu_Adaptive_2013,Gnanasambandam_IISW_2019,Gnanasambandam_TCI_HDR}.
\end{itemize}

For SPAD-QIS, the theoretical results are largely similar to those of CIS-QIS, although more attention are spent on modeling the dark current, timing uncertainty, and dead-time of the SPAD. In particular, \cite{Antolovic_DR_2018, Gupta_Passive_2019, Gupta_CVPR_2021} have some valuable results on the dynamic range of a one-bit SPAD. The quanta burst photography \cite{Gupta_QuantaBurst_2020} demonstrated the feasibility of using one-bit SPAD for high-speed imaging applications. On a different line of work in the context of non-line of sight imaging, analysis of SPAD has been reported in \cite{Shin_Nature_2016, Rapp_Unmixing_2017, Rapp_Deadtime_2019}.

The focus of this paper is on CIS-QIS. The main source of noise for CIS-QIS is the read noise. The dark current of these devices today is about 0.02 e-/s (e.g., the 2021 model by GigaJot Technology \cite{Ma_review_2022}). This allows us to skip the dark current in the theoretical derivations. However, it should be emphasized that the theoretical framework presented in this paper is generalizable to SPAD-QIS. This can be done by re-running the numerical simulation for a larger dark current.

\subsection{Main Results}
The key enabling tool of this paper is the recent analysis of the exposure-referred signal-to-noise ratio (SNR) in \cite{Chan_SNR_2022}. As shown in \cite{Chan_SNR_2022}, the SNR can be formally derived from statistical estimation theory, and several key equations can be written for the one-bit QIS. Building upon these results, three new theoretical findings are reported in this paper:

\begin{itemize}
\item Low-light: A CMOS image sensor performs worse at low-light because of the combination of a large read noise and the analog-to-digital conversion (ADC). One-bit quantization uses an extreme ADC (i.e., one bit), but it can truncate the tail of the Gaussian distribution of the read noise and offer a nearly noise-free signal. To achieve this, it is not necessary to push the read noise to 0 photoelectrons. A read noise of 0.15 photoelectrons is sufficient. However, if the exposure time is too short, the performance of the one-bit sensor will still degrade.
\item Frame rate: One-bit QIS can operate at a much higher frame rate because the one-bit signal is generated as soon as one photon is detected. Previous results of the Cramer Rao lower bound derived in \cite{Yang_Thesis_2012} showed that in the absence of noise, the frame rate should be made as high as possible. In the presence of read noise, however, this paper shows that an excessive frame rate will cause the per-frame signal-to-noise ratio to drop. The optimal frame rate is analytically derived. The equation matches with the numerical experiments.
\item Dynamic range: The benefit in the dynamic range comes in two pieces: the nonlinearity of the one-bit sensing and exposure bracketing. This paper shows that these two factors are completely independent of each other. The overall dynamic range is defined by the maximum and minimum integration time. New analytic expressions are derived.
\end{itemize}

The results provided in this paper are different from several previous work. Compared to \cite{Elgendy_Threshold_2018} which exclusively studies the threshold (in noise-free conditions), this paper addresses the questions of low-light, frame rate, and dynamic range. Compared to \cite{Gnanasambandam_TCI_HDR} which shows the dynamic range through numerical experiments, this paper shows analytic expressions and proves the decoupling of the nonlinearity and exposure bracketing. \cite{Chan_SNR_2022} gives a framework for the analysis, but this paper takes one step further to optimize the sensor parameters. The proof about the optimal frame rate in this paper is inspired by the derivations given in \cite{Chan_Density_2022}, but the motivations and conclusions are different.

\section{Mathematical Backgrounds}
One-bit QIS can be summarized by a simple Poisson-Gaussian model if high-order effects such as quantum efficiency, dark current, pixel non-uniformity, sensor gain, and threshold instability are not considered. In this case, the voltage generated at the floating diffusion of the photodiode is \cite{Fossum_Modeling_2013}
\begin{equation}
X = \text{Poisson}(\theta) + \text{Gaussian}(0,\sigma^2),
\label{eq: X}
\end{equation}
where $\theta$ is the underlying exposure of the scene integrated over the sensing area and the exposure time. The unit of $\theta$ is the average number of photoelectrons. The parameter $\sigma$ is the standard deviation of the read noise. Its unit is also the number of photoelectrons. Truncating the Poisson-Gaussian using a threshold $q$ yields a binary random variable
\begin{equation}
Y =
\begin{cases}
1, &\quad X \ge q,\\
0, &\quad X < q.
\end{cases}
\end{equation}

The probability density function $p_Y(y)$ (for $y = 0$ and $1$) is obtained by integrating the probability density function of $X$ from $q$ to $\infty$ \cite{Chan_SNR_2022}:
\begin{align*}
p_Y(1)
&= \int_{q}^{\infty} p_X(x) \; dx = \sum_{k=0}^{\infty} \frac{e^{-\theta}\theta^k}{k!} \Phi\left(\frac{k-q}{\sigma}\right),
\end{align*}
where $\Phi(z) = \int_{-\infty}^{z} \frac{1}{\sqrt{2\pi}}\exp\{-t^2/2\}dt$ is the cumulative distribution function of $\text{Gaussian}(0,1)$. Since $Y$ is binary, the probability of the other state is $p_Y(0) = 1-p_Y(1)$.

The bit density of the one-bit sensor is the average number of one's in a space-time volume. Mathematically, the bit-density is defined as \cite{Chan_Density_2022}
\begin{equation}
\mu \bydef \E[Y] = p_Y(1) = \sum_{k=0}^{\infty} \frac{e^{-\theta}\theta^k}{k!} \Phi\left(\frac{k-q}{\sigma}\right).
\label{eq: bit density}
\end{equation}
Plotting the bit density as a function of the exposure $\theta$ yields the classical D-logH curve.

An important tool to analyze the performance of a sensor is the signal-to-noise ratio (SNR). In the context of statistical estimation theory, the SNR of a pixel is defined as \cite{Yang_Thesis_2012}
\begin{equation}
\text{SNR}(\theta) \bydef \frac{\theta}{\sqrt{\E[(\thetahat(\mY) - \theta)^2]}},
\label{eq: SNR}
\end{equation}
where $\mY = \{Y_1,\ldots,Y_N\}$ is a sequence of i.i.d. observations of a pixel in the scene, and $\thetahat(\cdot)$ is an estimator. The estimator can be constructed from the maximum-likelihood (ML) estimation or another approach. For the special case where $\sigma = 0$ and $q = 0.5$, the ML estimator is as simple as $\thetahat(\mY) = -\log(1-\overline{Y})$ where $\overline{Y} = (Y_1+\ldots,Y_N)/N$ is the sufficient statistics of the measurement \cite{Chan_MDPI_2016}.

As discussed in \cite{Chan_SNR_2022}, the SNR defined in \eref{eq: SNR} can be approximated via the first-order Taylor expansion to obtain the \emph{exposure-referred SNR}:
\begin{equation}
\text{SNR}(\theta) \approx \sqrt{N} \cdot \frac{\theta}{\sqrt{\Var[Y]}}\cdot \frac{d\mu}{d\theta},
\label{eq: SNR exp}
\end{equation}
where $\mu \bydef \E[Y]$ and $Y$ is any of $Y_1,\ldots,Y_N$. The exposure-referred SNR is identical to the asymptotic SNR defined through the Fisher Information by Yang et al. \cite{Yang_Thesis_2012} of which the equivalence can be found in Elgendy and Chan \cite{Elgendy_Threshold_2018}. One way to think about the exposure-referred SNR is that the derivative $d\mu/d\theta$ plays the role of a transfer function that provides a ``gain'' to the variance $\Var[Y]$. This transfer function $d\mu/d\theta$ amplifies or attenuates the variance caused by the nonlinearity of the sensor. In the sensor community, the exposure-referred SNR is also known as the input-referred SNR \cite{EMVA_2010}.

On a computer, the exposure-referred SNR can be implemented easily for one-bit sensors. The intuition is that in the case of a one-bit truncated Poisson-Gaussian, the mean of the random variable $Y$ is $\mu = \E[Y] = p_Y(1)$. Since $Y$ is binary, it follows from the Bernoulli statistics that the variance is $\Var[Y] = \mu(1-\mu)$. Therefore, for one-bit sensors
\begin{align}
\text{SNR}_{\text{QIS}}(\theta) = \sqrt{N} \cdot \frac{\theta}{\sqrt{\mu(1-\mu)}}\cdot\frac{d\mu}{d\theta},
\end{align}
where $\mu = \E[Y]$ is the bit-density defined in \eref{eq: bit density}. Once $\mu$ and $\Var[Y]$ are determined, the derivative $d\mu/d\theta$ can be approximately determined using the finite difference method \cite{Chan_SNR_2022}. Since $\mu = \E[Y]$ is a smooth function, any standard finite-difference operation would suffice to provide a good estimate of $d\mu/d\theta$. The MATLAB code that generates the exposure-referred SNR for a one-bit signal is shown in \cite[Appendix]{Chan_SNR_2022}.

For conventional CMOS image sensors (CIS) with a full well capacity FWC, the exposure-referred SNR can be approximated by the standard linear model:
\begin{align}
\text{SNR}_{\text{CIS}}(\theta)
=
\begin{cases}
\frac{\theta}{\sqrt{\theta + \sigma^2}}, &\qquad \theta < \text{FWC},\\
0, &\qquad \theta \ge \text{FWC}.
\end{cases}
\end{align}
Notice that the integer $N$ is not included in this equation because the CIS only takes one frame whereas a QIS takes $N$ binary frames.

\section{Benefit 1: Low-light}
One of the biggest motivations in the original QIS proposal is to count photons in low-light conditions. Therefore, the first question to be answered is the potential benefits of the one-bit QIS compared to a CMOS image sensor (CIS) at low-light.

\subsection{Perfect Sensor $\not\Rightarrow$ Noise-Free}
Before discussing the sensor's performance, one clarification is needed for readers who are new to image sensors: A perfect sensor does not produce a noise-free image. The notion of ``low-light'' refers to the condition at which $\theta$ is so small that the Poisson random variable dominates the signal. Since the Poisson process is driven by the photoelectric conversion, the randomness is present even if the sensor is free from any circuit artifacts.

\begin{figure}[h]
\centering
\begin{tabular}{ccc}
\hspace{-1ex}\includegraphics[width=0.32\linewidth]{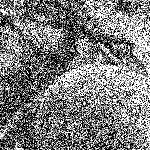}&
\hspace{-2ex}\includegraphics[width=0.32\linewidth]{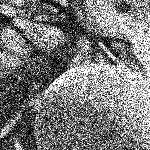}&
\hspace{-2ex}\includegraphics[width=0.32\linewidth]{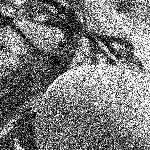}\\
\hspace{-1ex}(a) 14-bit CIS & \hspace{-2ex} (b) 14-bit CIS & \hspace{-2ex} (c) perfect sensor\\
$\sigma = 2$  & $\sigma = 0.15$ & $\sigma = 0$
\end{tabular}
\vspace{-2ex}
\caption{\textbf{A perfect sensor $\not=$ noise-free}. This example considers the case where $\theta \in [0,5]$. For a conventional CMOS image sensor (CIS) with a read noise of $\sigma = 2$, the image is noisy. If the sensor uses a lower read noise of $\sigma = 0.15$, the image quality is improved but it is still limited. Even if the sensor is a hypothetical ``perfect sensor'' with zero read noise, there is still noise caused by the Poisson random variable.}
\label{fig: perfect sensor}
\vspace{-2ex}
\end{figure}

Consider a simple experiment. Suppose that there is a 14-bit CIS with a read noise of $\sigma = 2$. Compare this sensor with another 14-bit CIS with a read noise of $\sigma = 0.15$ and a 14-bit ideal image sensor with zero read noise. The simulated results shown in \fref{fig: perfect sensor} are based on the Poisson-Gaussian distribution. As shown, in low-light conditions, even a perfect sensor is limited by the Poisson noise.

The message here is that when comparing one-bit QIS with CIS, the question is not about whether the image can be free of \emph{any} noise. Instead, the objective is to check whether the captured signal has faithfully reported the Poisson random variable. This should be kept in mind because all images shown in this paper are noisy.

\subsection{One-bit Quantization Truncates Read Noise}
The low-light performance of the one-bit QIS comes in two parts. The first one is the small read noise enabled by the pump-gate technology in QIS jot design and the advanced fabrication techniques \cite{Ma_Pump_2015}. As reported in \cite{Ma_019e_2021}, the latest QIS has achieved a read noise of 0.19 electrons at room temperature. Compared to a CIS where the read noise is several photons, this is significant improvement.

\begin{figure}[ht]
\centering
\includegraphics[width=0.95\linewidth]{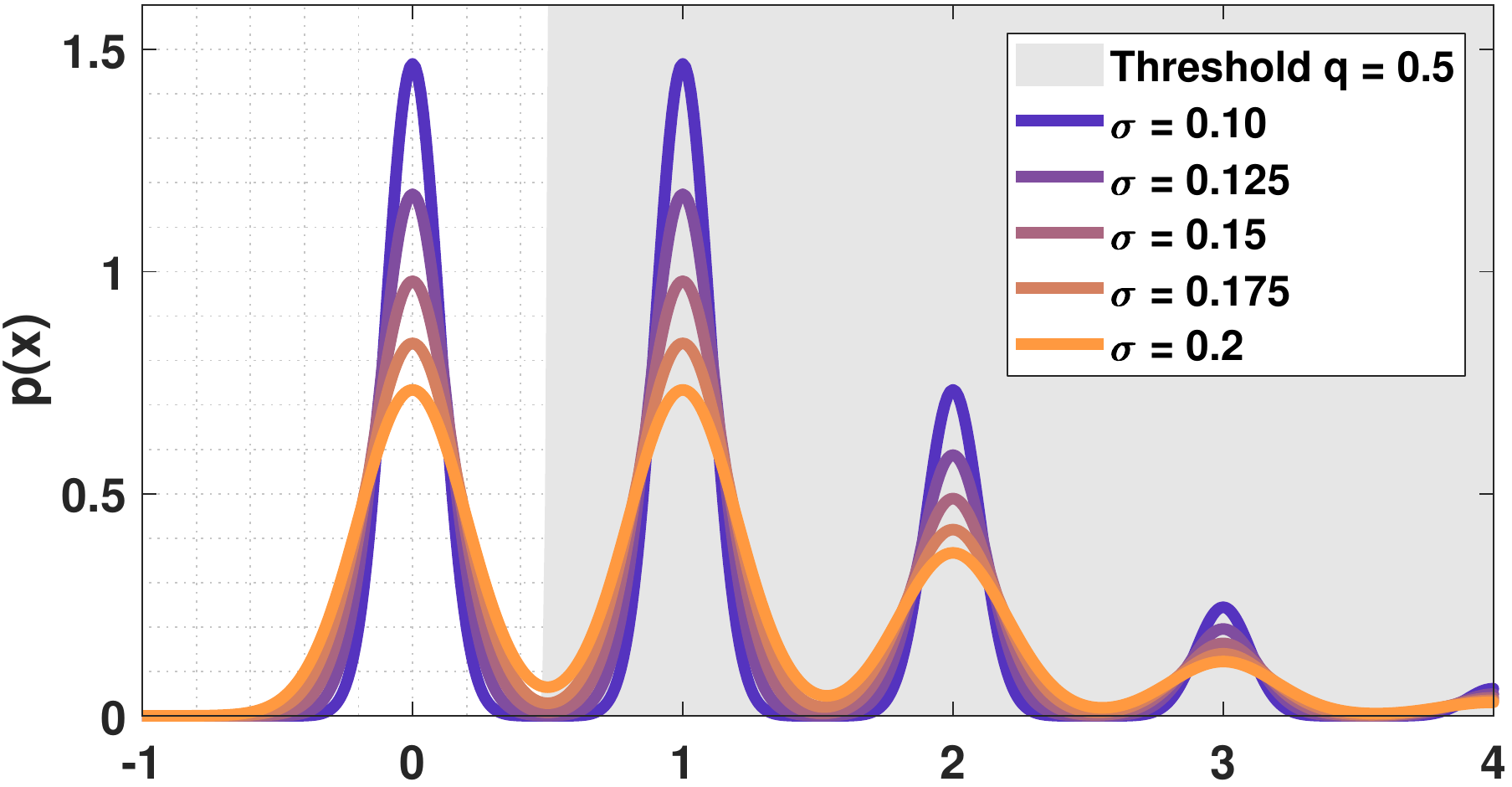}
\caption{\textbf{Poisson-Gaussian distribution when $\theta = 1$}. Notice that for $\sigma \le 0.15$, the Gaussians do not interfere with other Gaussians.}
\label{fig: 1b read noise level a}
\end{figure}

The second source of the improvement is the one-bit quantization commented by Fossum \cite{Fossum_Modeling_2013}. It is easy to show that the empirical bit density is the sufficient statistic for the underlying parameter $\theta$. When the Poisson-Gaussian distribution is truncated to a binary bit, the error in the bits (i.e., ``1'' becomes ``0'' or ``0'' becomes ``1'') is the area of the curve covered by the tail part of the Gaussian, as shown in \fref{fig: 1b read noise level a}. If $\theta = 1$ and threshold $q = 0.5$, the necessary and sufficient condition for an error-free bit density is that
\begin{align}
\Phi\left(\frac{q}{\sigma}\right) = \underset{\text{\scriptsize{area under a Gaussian distribution from -$\infty$ to $q$}}}{\underbrace{\int_{-\infty}^{q} \frac{1}{\sqrt{2\pi\sigma^2}} e^{-\frac{x^2}{2\sigma^2}} \; dx}} \ge 1-\delta,
\label{eq: read noise bound}
\end{align}
where $\delta$ is the desired level of tolerance. Rearranging the terms gives $\sigma \le 0.5/\Phi^{-1}(1-\delta)$. Thus, to achieve a specific tolerance level $\delta$, the read noise $\sigma$ has to be no greater than the bound shown in \eref{eq: read noise bound}. As shown in \fref{fig: 1b read noise level a}, $\sigma \le 0.15$ is typically sufficient to preserve the bit density.

\begin{figure}[ht]
\centering
\includegraphics[width=0.95\linewidth]{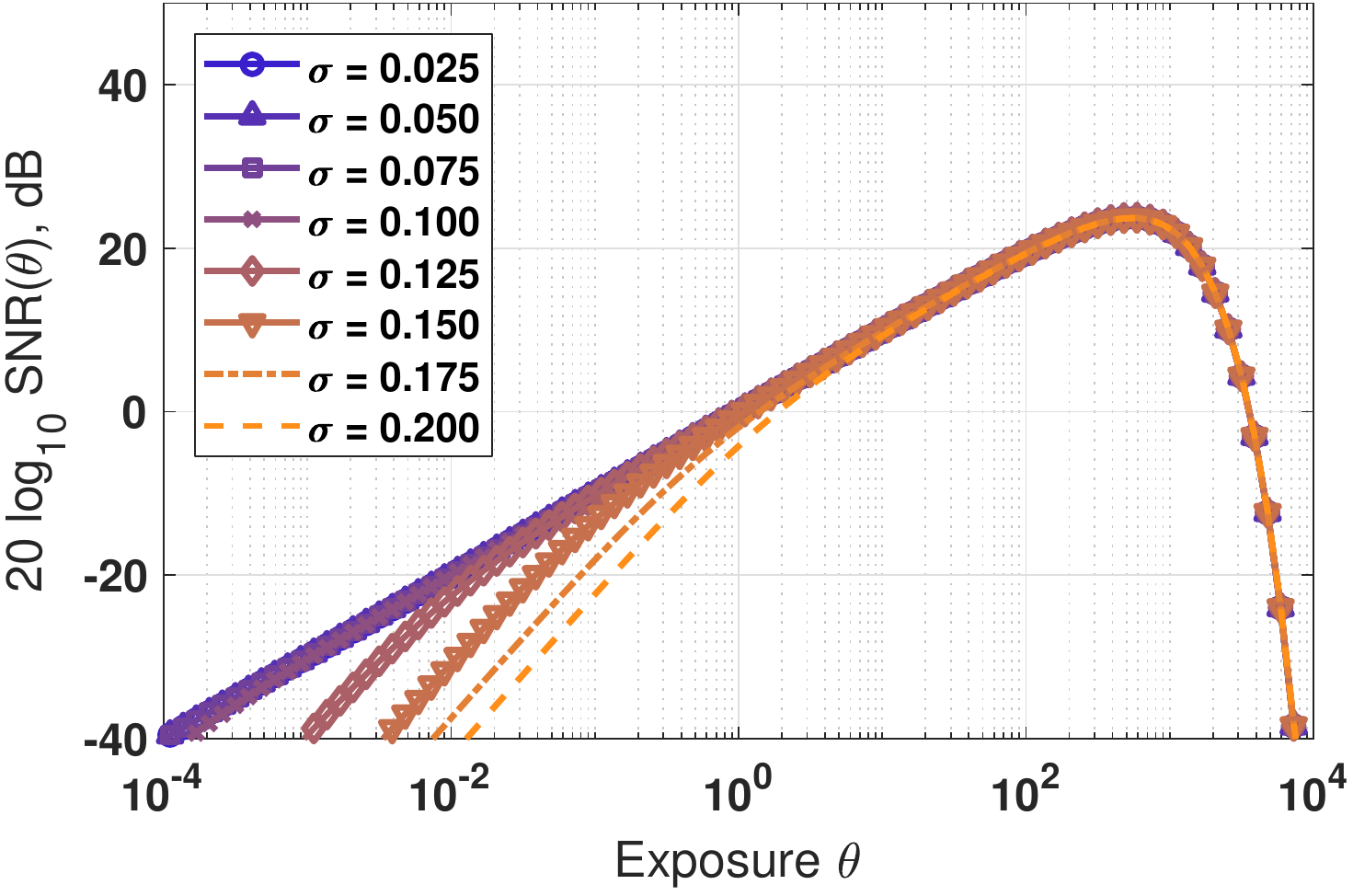}
\caption{\textbf{One-bit QIS does not need $\sigma = 0$}. The SNR of $N = 256$ frames of binary bits truncated at $q = 0.5$. Notice that the SNR degrades only at very low-light.}
\label{fig: 1b read noise level b}
\vspace{-1ex}
\end{figure}

However, the analysis based on the bit density (as in \cite{Fossum_Modeling_2013}) does not cover the complete story. One thing to note is that \fref{fig: 1b read noise level a} is plotted for $\theta = 1$. When $\theta < 1$, since the probability of the Poisson random variable at $X = 0$ is higher, the read noise needs to be smaller in order to preserve the bit density. A more consistent analysis can be conducted by using the exposure-referred SNR defined in \eref{eq: SNR exp}. Following the derivations in \cite{Chan_SNR_2022}, the SNR for the one-bit signal has a closed-form expression which can then be plotted in \fref{fig: 1b read noise level b}. For this particular plot, the number of frames is set as $N = 256$, and the exposure $\theta$ is ranged from $10^{-4}$ to $10^4$. Across the range of exposure levels, the SNR shows a different behavior as the read noise becomes stronger. In particular, when $\theta = 10^{-4}$, the read noise has to be as small as $\sigma \le 0.1$ to ensure a consistent SNR.

The new finding raises a question: does it mean that the read noise has to be smaller than what people usually think ($\sigma = 0.15$)? The answer depends on the integration time. The $x$-axis of the SNR plot in \fref{fig: 1b read noise level b} is the average number of photoelectrons. If the scene is dark (aka low-light), the integration time can be set longer to maintain the average number of photoelectrons. As long as the sensor operates approximately at $\theta = 1$, a read noise of $\sigma = 0.15$ is sufficient to guarantee that the binary bits have a negligible probability of error. A \emph{zero} read noise is not necessary.

\subsection{Dividing the Exposure into Many One-bit Frames}
The results in \fref{fig: 1b read noise level b} are based on using $N = 256$ binary frames. To reach a definitive conclusion about the benefit of one-bit QIS over a multi-bit CIS, it is necessary to find an \emph{equivalent} configuration between the two. The following new analysis is developed to address this question.

Consider a total exposure of $\theta$. This $\theta$ is seen by both the one-bit QIS and a multi-bit CIS. For CIS, one image is taken, whereas for one-bit QIS, a total of $N$ frames are taken. The number $N$ has to match with the bit-depth of the CIS. For example, if the CIS has a full-well capacity of $\text{FWC} = 4096$ electrons, then the equivalent QIS can use $N = 4096$ binary frames. That is, assuming a uniform division of the exposure across the $N$ frames, the exposure seen during each QIS capture is $\theta/N$.

Under this condition, the mean of one binary random variable $Y$ is
\begin{equation}
\mu_N \bydef \E[Y] = \sum_{k=0}^{\infty} \frac{e^{-\tfrac{\theta}{N}}(\tfrac{\theta}{N})^k}{k!} \Phi\left(\frac{k-q}{\sigma}\right),
\label{eq: mu_N}
\end{equation}
and hence the variance is $\Var[Y] = \mu_N(1-\mu_N)$. When averaged over $N$ frames, the overall SNR becomes
\begin{equation}
\text{SNR}_{\text{QIS}, N}(\theta) = \sqrt{N} \cdot \frac{\theta}{\sqrt{\mu_N(1-\mu_N)}} \cdot \frac{d\mu_N}{d\theta},
\label{eq: SNR_N}
\end{equation}
where the subscript $N$ emphasizes the $\mu$ depends on $N$.

For CIS, the SNR is based on one frame and therefore it can be approximated by
\begin{equation}
\text{SNR}_{\text{CIS}, 1}(\theta) = \frac{\theta}{\sqrt{\theta + \sigma^2}}, \;\; \text{for} \;\; \theta \le \text{FWC},
\end{equation}
and $\text{SNR}_{\text{CIS}, 1}(\theta) = 0$ for $\theta > \text{FWC}$.

The SNRs of the QIS and the CIS for different $N$ are shown in \fref{fig: low-light}. In this plot, the FWC of the CIS is set as $\text{FWC} = 4096$. The read noise of CIS is $\sigma = 2$ and the read noise of QIS is $\sigma = 0.1$. The threshold is $q = 0.5$. Two observations can be drawn from this plot:
\begin{figure}[h]
\centering
\includegraphics[width=\linewidth]{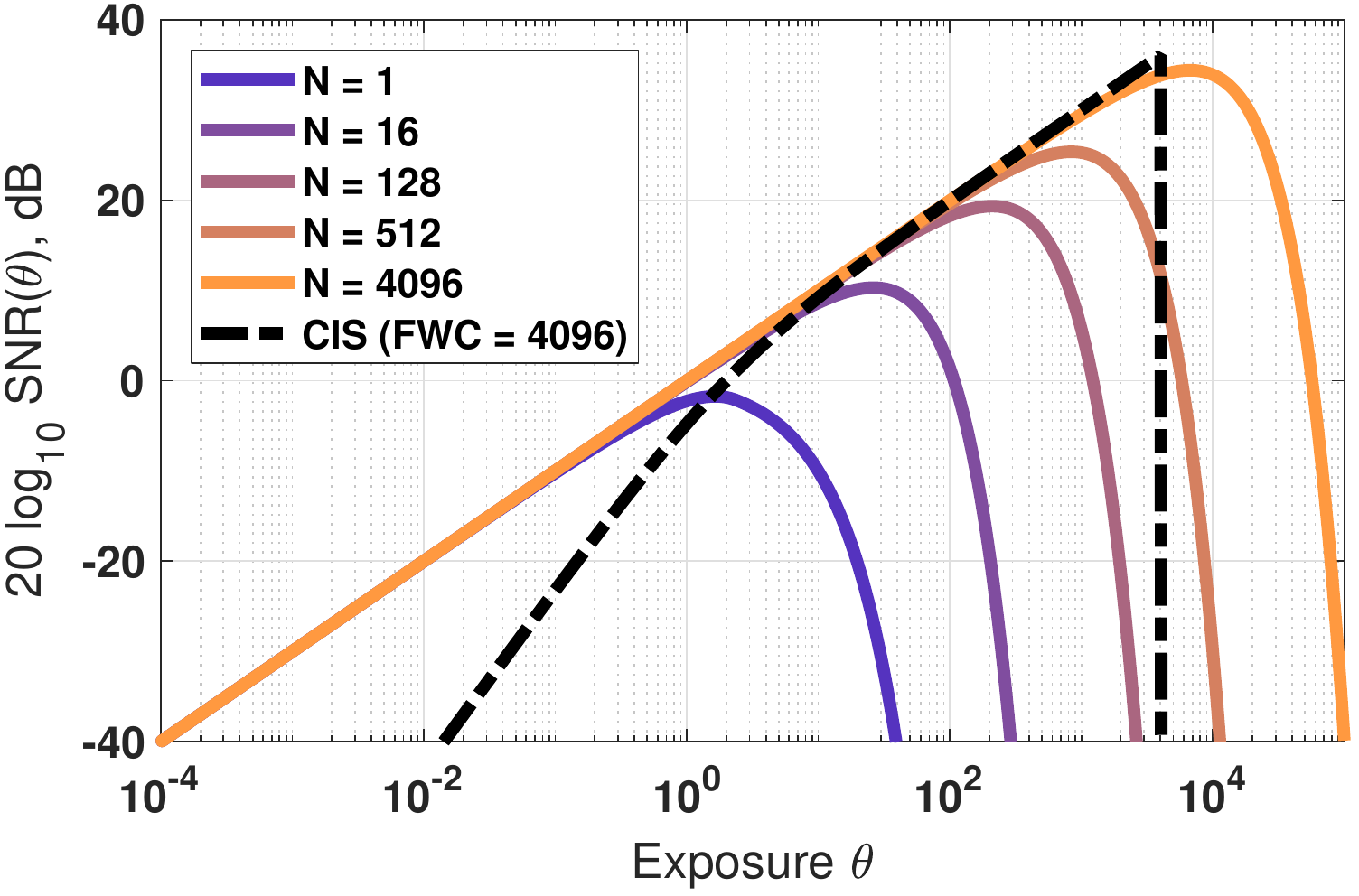}
\vspace{-4ex}
\caption{\textbf{QIS vs CIS, assuming $\sigma = 0.1$}. The QIS is assumed to have a read noise of $\sigma = 0.1$ whereas the CIS is assumed to have a read noise of $\sigma = 2$. For fair comparisons, both sensors has the identical total exposure. QIS divides the exposure into $N$ segments where each uses ($1/N$)th of the total exposure.}
\label{fig: low-light}
\vspace{-2ex}
\end{figure}

\begin{itemize}
\item \emph{QIS has better performance at low-light}. For a small $\theta$ such as $\theta \le 10^0$, the SNR of the QIS is consistently better than the CIS. This is attributed to the noise rejection capability of the one-bit quantization.
\item \emph{QIS has equal performance at high-light}. For large $\theta$ such as $\theta \ge 10^1$, QIS performs better when a sufficiently large $N$ is used. Intuitively, if $\theta$ is large, most of the bits will be 1 if the per-frame integration time is too long. By increasing $N$ (and hence a shorter integration time per frame), the bit density will be more difficult to saturate. Thus, the SNR of the QIS improves. However, if $\theta$ is large, CIS does not suffer from the read noise anymore. Its performance is hence on par with the QIS.
\item \emph{Peak SNRs are similar}. Assuming that CIS has a full-well capacity of $N$ whereas QIS uses $N$ binary frames, the peak SNRs of the QIS and CIS are comparable.
\end{itemize}

\begin{figure}[h]
\centering
\includegraphics[width=\linewidth]{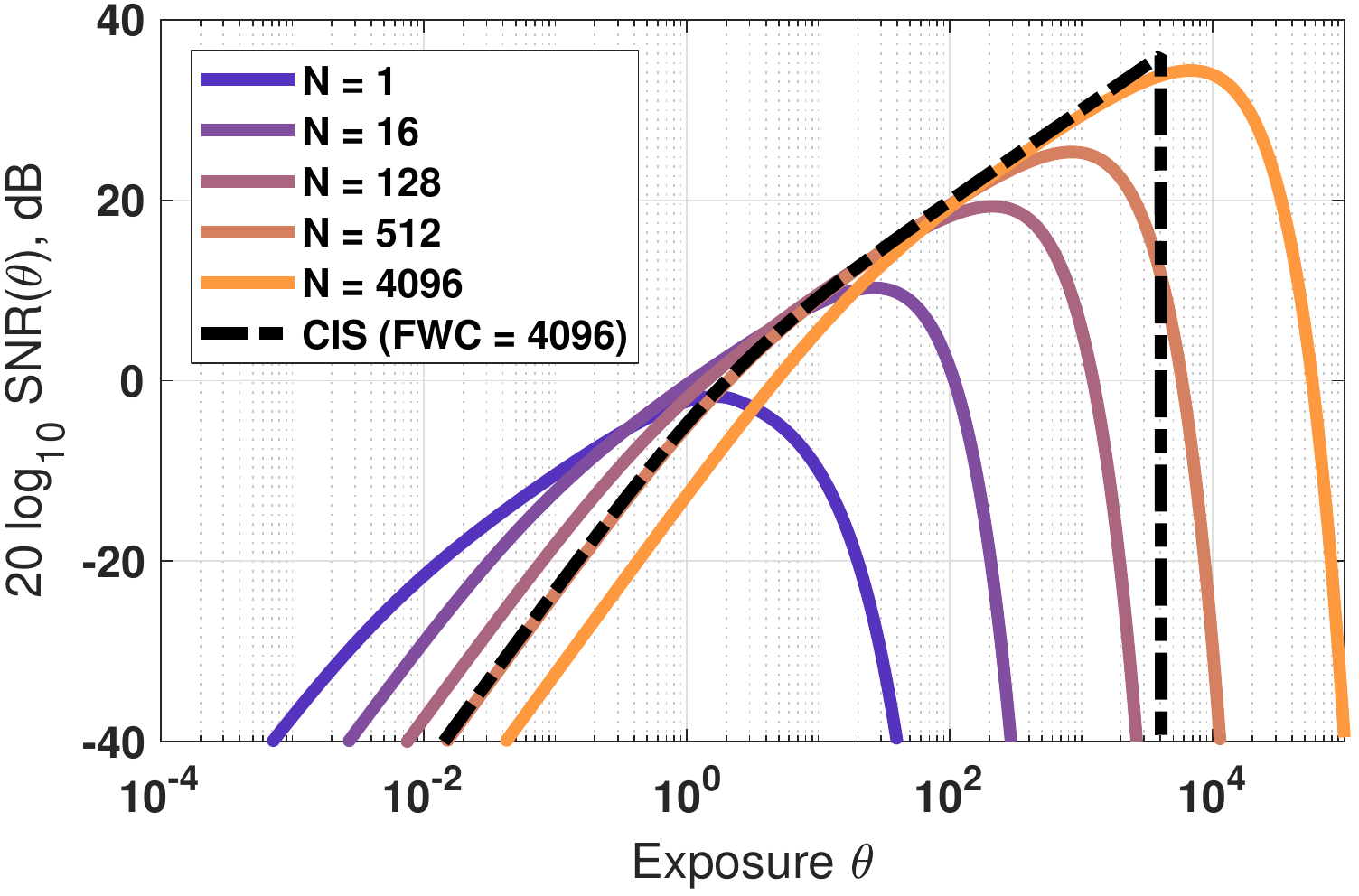}
\vspace{-4ex}
\caption{\textbf{QIS vs CIS, assuming $\sigma = 0.19$}. The configuration of this plot is the same as \fref{fig: low-light}, except that the read noise of QIS is $\sigma = 0.19$. Because of the increased read noise, using a high frame rate $N = 4096$ will have a worse performance. One solution to mitigate this is to use an exposure protocol where within the total integration time, QIS divides it into a collection of variable-exposures so that the low-light is taken care by small $N$ whereas the high-light is taken care by the large $N$.}
\label{fig: low-light 2}
\end{figure}

When read noise QIS is at a higher level, such as $\sigma = 0.19$, the SNR will behave as shown in \fref{fig: low-light 2}. The configurations used to generate this theoretical plot is identical to \fref{fig: low-light}, except that the read noise of QIS is $\sigma = 0.19$. Because of the increased read noise, the performance of QIS becomes worse when $N$ is large. When the integration time per capture is very short, the Poisson component $\text{Poisson}(\theta)$ will be weaker than the read noise $\text{Gaussian}(0,\sigma^2)$.

The implication of this set of experiments is that the one-bit QIS is not always better than the multi-bit CIS. For high-light regions, QIS performs equally well with CIS. For low-light, QIS requires a longer integration time (and hence a smaller $N$) to outperform CIS. If the per-frame integration time is not configured properly, CIS can still perform better.

\subsection{CIS Suffers from Analog-Digital Conversion at Low-Light}
When comparing CIS's and QIS's performance at low-light, one more factor needs to considered. Both one-bit QIS and multi-bit CIS use the analog-to-digital converter (ADC), but for QIS, the ADC is one-bit. If there are $N$ binary frames, the effective number of levels after summing these $N$ frames will be $N$. For a multi-bit CIS, if the full-well capacity is FWC and it uses an $L$-bit ADC (so that there are $2^L$ levels), the quantized measurement will be
\begin{align*}
Y = \text{ADC}(X)
=
\text{round}\left( \frac{X}{\text{FWC}} \cdot 2^L \right) \cdot \frac{\text{FWC}}{2^L},
\end{align*}
where $X = \text{Poisson}(\theta) + \text{Gaussian}(0,\sigma^2)$ is the analog voltage. In other words, the ADC divides the signal range from 0 to FWC into $2^L-1$ equally spaced segments. Within each segment the analog voltage is integrated to produce a digital number. Therefore, if the scene is dark so that $X$ is small, the quantized measurement $Y$ will be truncated to zero.

\begin{figure}[ht]
\centering
\begin{tabular}{cc}
\hspace{-1ex}\includegraphics[width=0.49\linewidth]{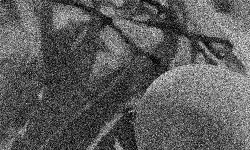}&
\hspace{-2ex}\includegraphics[width=0.49\linewidth]{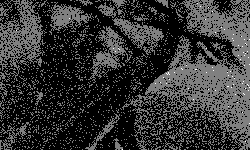}\\
(a) one-bit QIS, 16 frames & (b) 8-bit CIS, 1 frame
\end{tabular}
\vspace{-2ex}
\caption{\textbf{Impact of the analog-digital conversion}. The exposure lies in the range $\theta \in [0,16]$. The QIS model uses a read noise of $\sigma = 0.19$, $N = 16$ binary frames where each frame sees an exposure $\theta/N$. The CIS model uses a read noise of $\sigma = 2$, a full-well capacity of FWC = 4000, and a 8-bit ADC with $2^{8} = 256$ levels.}
\label{fig: gain and ADC}
\end{figure}

The impact of the ADC can be visualized in \fref{fig: gain and ADC}. In this example, the underlying exposure is assumed to lie in the range of $\theta \in [0,16]$. The QIS is assumed to have a read noise of $\sigma = 0.19$. A total number of $N = 16$ frames are used to construct the one-bit measurements. For CIS, the read noise is $\sigma = 2$. The full well capacity is $\text{FWC} = 4000$, and the bit-depth is $L = 8$ so that the ADC has $2^L = 256$ levels. The results shown in \fref{fig: gain and ADC} demonstrate the quality of the images under this configuration. Although one frame of the 8-bit CIS has 256 levels while the sum of $N$ one-bit QIS frame only has $16$ levels, the quantization of the former is too coarse to keep the weak signal under low light.

To summarize this section, the low-light performance of QIS and CIS depends on multiple factors: $\theta$, $\sigma$, $N$, and ADC. There are configurations where CIS performs better.

\section{Benefit 2: Frame Rate}
A one-bit sensor can operate at a much higher speed than a conventional camera. Prototypes of SPAD-QIS have achieved more than 16k frames/second \cite{Dutton_QVGA_2015} to 150k frames/second \cite{Burri_Monolithic_2018}. Thus, whether the sensor \emph{can} operate at high speed is no longer a question. A more meaningful question here is whether it \emph{should} be operating at such a high speed, or should the speed be scaled with respect to the brightness of the scene? If so, what is the relationship?

\subsection{Higher Frame Rate $\not\Rightarrow$ ``Better''}
\begin{figure}[h]
\vspace{-1ex}
\centering
\begin{tabular}{ccc}
\hspace{-1ex}\includegraphics[width=0.3\linewidth]{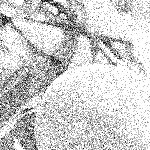}&
\hspace{-2ex}\includegraphics[width=0.3\linewidth]{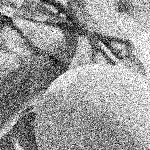}&
\hspace{-2ex}\includegraphics[width=0.3\linewidth]{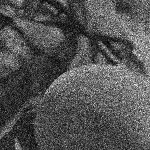}\\
$N = 4$ & $N = 8$ & $N = 256$\\
\hspace{-1ex}\includegraphics[width=0.3\linewidth]{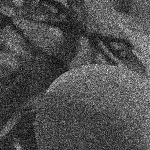}&
\hspace{-2ex}\includegraphics[width=0.3\linewidth]{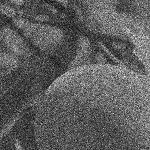}&
\hspace{-2ex}\includegraphics[width=0.3\linewidth]{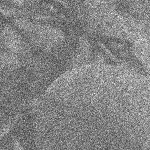}\\
$N = 512$ & $N = 1024$ & $N = 4096$\\
\end{tabular}
\vspace{-1ex}
\caption{\textbf{Impact of frame rate.} The underlying scene has an exposure $\theta \in [0,16]$. The sensor captures the scene using $N$ frames where each frame is exposed to $\theta/N$. The read noise is assumed to be $\sigma = 0.2$.}
\label{fig: frame rate images}
\end{figure}

The main conclusion about the frame rate is that the frame rate can be infinity only when there is no read noise. \fref{fig: frame rate images} demonstrates the intuition and the subsequent subsections will prove this statement. \fref{fig: frame rate images} considers a scene where the underlying exposure is in the range of $[0,16]$. A total of $N$ one-bit frames are drawn where each frame has an exposure $\theta/N$. As shown in the figure, if the frame rate is too low ($N$ is small), the observed image quality is poor because there is not enough frames to compensate for the one-bit quantization. But if the frame rate is too high ($N$ is large), the image quality will degrade because the per-frame read noise adds up.

\begin{figure}[h]
\centering
\includegraphics[width=\linewidth]{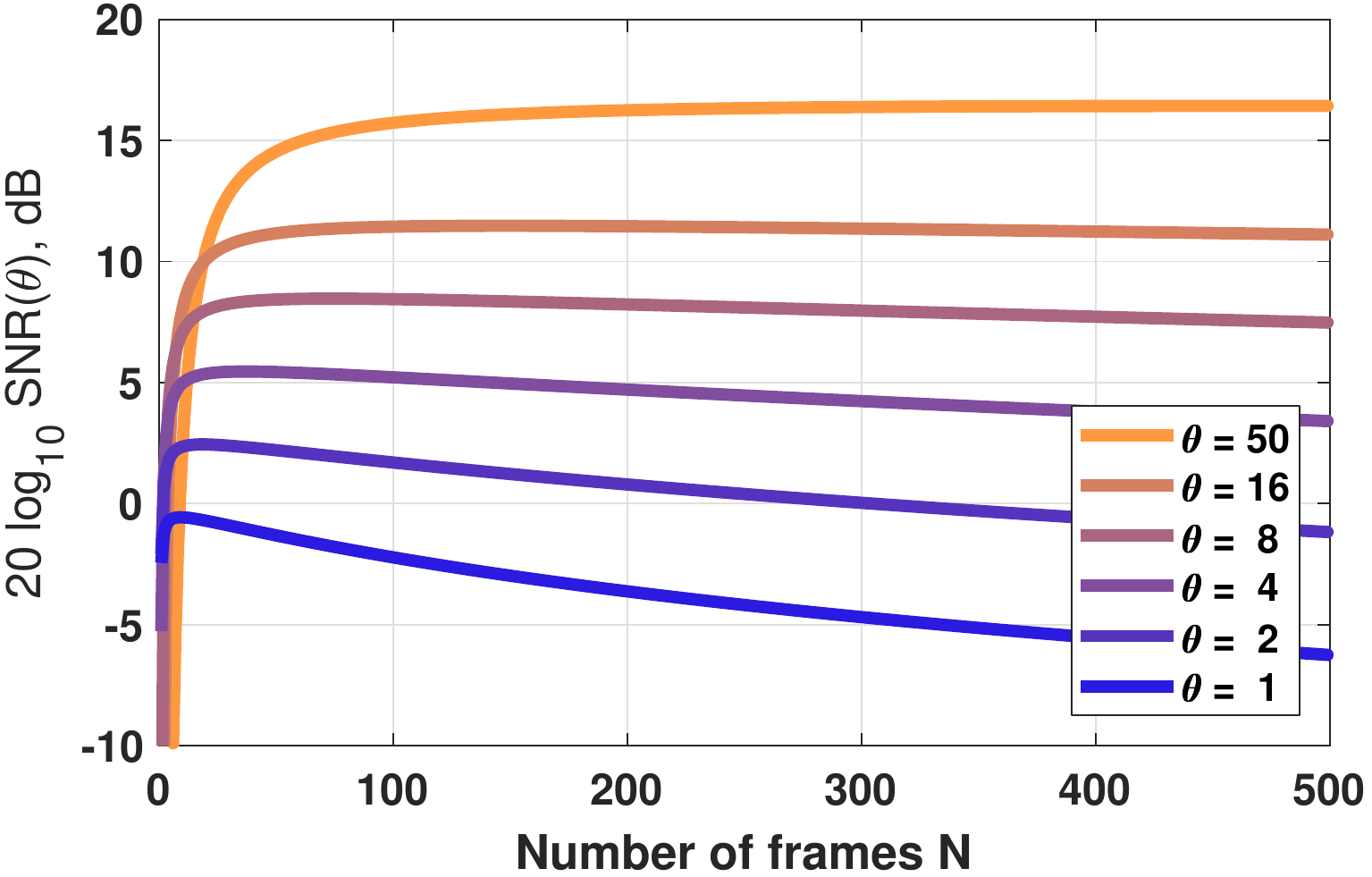}
\vspace{-4ex}
\caption{\textbf{One-bit sensors cannot use too many frames.} The model used in generating this plot is the Poisson-Gaussian equation where the total exposure $\theta$ is divided by the number of frames $N$, i.e., $\text{Poisson}(\theta/N) + \text{Gaussian}(0,\sigma^2)$, for $N$ times. In this experiment, the read noise is $\sigma = 0.2$. The plot shows the SNR as a function of $N$. For every $\theta$, there exists an optimal $N$ where the SNR is maximized. Thus the frame rate cannot be arbitrarily high.}
\label{fig: Number of Frames}
\end{figure}

To further illustrate the impact of the frame rate to the SNR, a plot showing how the SNR would change as a function of $N$ is drawn in \fref{fig: Number of Frames}, for different values of $\theta$. This is a new result in the literature. What this plot shows is that if $N$ is large, SNR will degrade. Moreover, for every $\theta$, there is an optimal $N$ where the SNR is maximized. This optimal $N$ can be theoretically predicted.

\subsection{Optimal Frame Rate: Noise-Free Case}
With the intuitions discussed, this sub-section presents the theoretical justifications. The objective here is to find an $N$ that maximizes the SNR for any given $\theta$ and $\sigma$. However, considering the Poisson-Gaussian distribution, such a derivation would require some less-known mathematical tools. One of them is the Lambert-W function.
\begin{definition}[Lambert-W function, \cite{LambertW}]
The Lambert-W function $W(\cdot)$ is the inverse mapping of the function $f(x) = xe^{x}$. That is, $W(c)$ is the solution to the problem
\begin{equation}
xe^{x} = c
\end{equation}
and this holds for any $c \ge -1/e$.
\end{definition}

The Lambert-W function is used in a variety of thermodynamics and quantum mechanics problems. It cannot be expressed in terms of elementary functions. As a quick example, the solution of $xe^{x} = 5$ is $x = W(5)$.\footnote{The Lambert-W function has two roots, namely $W_0(c)$ for $c\ge 0$ and $W_{-1}(c)$ for $-1/e \le c < 0$. This paper mainly focuses on the negative part.}

With the Lambert-W function, one can analytically derive the maximum SNR. The following theorem is a new result about the optimal $\theta$ and $N$ for the case $\sigma = 0$.
\begin{theorem}
\label{thm: max SNR, theta}
Consider the SNR of a one-bit signal defined in \eref{eq: mu_N} and \eref{eq: SNR_N}, assuming $\sigma=0$. For fixed number of frames $N$, the maximizer $\theta^*$ of the SNR is
\begin{align}
\theta^*
&= \argmax{\theta} \;\; \text{SNR}_{\text{QIS},N}(\theta) = N [ 2 + W(-2 e^{-2})],
\end{align}
where $W(\cdot)$ is the Lambert-W function. For a fixed exposure $\theta$, the optimal $N^*$ that maximizes the SNR is
\begin{equation}
N^* = \argmax{N} \;\; \text{SNR}_{\text{QIS},N}(\theta) = \infty.
\end{equation}
\end{theorem}

\begin{proof}
Recall the SNR defined in \eref{eq: SNR_N}. When $\sigma = 0$, the mean $\mu_N$ and its derivative are
\begin{equation*}
\mu_N = 1-e^{-\frac{\theta}{N}}, \quad \frac{d\mu_N}{d\theta} = \frac{1}{N}e^{-\frac{\theta}{N}}.
\end{equation*}
Therefore, the SNR is
\begin{align*}
&\text{SNR}_{\text{QIS},N}(\theta) = \sqrt{N} \cdot \frac{\theta}{\sqrt{\mu_N(1-\mu_N)}}\cdot\frac{d\mu_N}{d\theta} \\
&= \frac{\sqrt{N}\theta}{\sqrt{(1-e^{-\frac{\theta}{N}})(e^{-\frac{\theta}{N}})}} \cdot \frac{1}{N}e^{-\frac{\theta}{N}} = \frac{\theta}{\sqrt{N} \sqrt{e^{\frac{\theta}{N}}-1}}.
\end{align*}
Since the goal is to maximize the SNR, any monotonically increasing function such as a log would not alter the maximizer. Thus, we consider
\begin{align}
&\log \text{SNR}_{\text{QIS},N} (\theta) \notag \\
&\qquad\qquad = - \frac{1}{2}\log N  + \log \theta - \frac{1}{2}\log\left(e^{\frac{\theta}{N}}-1 \right).
\label{eq: SNR QIS N}
\end{align}
The derivative of the log SNR with respect to $\theta$ is
\begin{align*}
\frac{d}{d\theta} \log \text{SNR}_{\text{QIS},N}(\theta) = \frac{1}{\theta} - \frac{e^{\frac{\theta}{N}}}{2N(e^\frac{\theta}{N}-1)} = 0.
\end{align*}
Rearranging the terms yields $\left(\frac{\theta}{N}-2\right)e^{\frac{\theta}{N}-2} = -2e^{-2}$, and hence the solution is $\theta^* = N\left(2+W(-2e^{-2})\right)$.

To prove the second result, one can take derivative of the SNR with respect to $N$. This will give us
\begin{align*}
\frac{d}{dN}\log \text{SNR}_{\text{QIS},N}(\theta) = \frac{-1}{2N} + \frac{e^{\frac{\theta}{N}}\theta}{2N^2(e^{\frac{\theta}{N}}-1)}.
\end{align*}
Setting it to zero and re-arranging the terms will give us $\left(\frac{\theta}{N}-1\right)e^{\frac{\theta}{N}-1} = -e^{-1}$. Therefore, the solution is
\begin{equation*}
N = \frac{\theta}{1+W(-e^{-1})}.
\end{equation*}
However, since $W(-e^{-1}) = -1$, it follows that $N = \infty$.
\end{proof}

The result of Theorem~\ref{thm: max SNR, theta} explains several important points:
\begin{itemize}
\item Theorem~\ref{thm: max SNR, theta} provides the first analytic expressions of $\theta^*$ and $N^*$ in the one-bit QIS literature. While numerical computation is still needed to evaluate the peak SNRs, it no longer requires an optimization toolbox to iteratively search for the maximum point.
\item The optimal exposure that maximizes the SNR is $\theta^* = N(2+W(-2e^{-2}))$. Except for the factor $N$, the rest of the equation is a constant which does not depend on the sensor parameters.
\item The optimal number of frames is $N = \infty$ when the read noise is zero. This is because when there is no read noise, every frame will be a faithful measurement of the scene intensity. The shorter the exposure is (by using a large $N$), the less likely a frame will be saturated. In the extreme, when $N$ is approaching infinity, each photon is guaranteed to be perfectly measured by one of the frames. There is no harm in using more frames. As predicted by the theory, $N = \infty$.
\item The theoretical result helps explain the dynamic range (which will be discussed in more detail in the next section.) In the absence of noise, the maximizer $\theta^*$ occurs at $N(W(-2e^{-2})+2) \approx 1.5936N$. If $N$ is the full-well capacity FWC, then $\theta^* = 1.5936 \times \text{FWC}$. So if the CIS saturates at the FWC, one-bit sensors do not saturate until $1.59 \times \text{FWC}$. This is about 50\% more dynamic range, and if exposure bracketing is considered, the dynamic range will be even wider.
\end{itemize}

\subsection{Optimal Frame Rate: Noisy Case}
The above analysis assumes a zero read noise. In the presence of read noise, some modifications can lead to the desired result. Consider the Poisson-Gaussian distribution:
\begin{equation}
p_Y(1) = \sum_{k=0}^{\infty}
\underset{\calP_{\theta}(k)}{\underbrace{\frac{e^{-\frac{\theta}{N}}\left(\frac{\theta}{N}\right)^k}{k!}}}
\underset{\calG_{\sigma}(k)}{\underbrace{\Phi\left(\frac{k-q}{\sigma}\right)}}.
\label{eq: p_Y(1) for optimal N}
\end{equation}
The following lemma shows the analytic expression of the SNR, in terms of the amount of read noise. This is a new result which generalizes \cite{Chan_SNR_2022}, and is based on a similar technique developed in \cite{Chan_Density_2022}.

\begin{lemma}
Consider a one-bit sensor with a read noise $\sigma > 0$. Let $\omega = 1-\Phi(-0.5/\sigma)$ where $\Phi(\cdot)$ is the cumulative distribution function of the standard Gaussian. Then, for small $\sigma$ (typically $\sigma \le 0.2$), the SNR can be approximated by
\begin{equation}
\text{SNR}_{\text{QIS},N}(\theta) = \frac{\theta}{\sqrt{N}}\cdot \sqrt{\frac{1}{\frac{1}{\omega}e^{\frac{\theta}{N}}-1 }}.
\end{equation}
\end{lemma}

\begin{proof}
The difference between the noise-free case and the noisy case is the presence of the Gaussian part $\calG_{\sigma}(k)$ in \eref{eq: p_Y(1) for optimal N}. When $\sigma = 0$, $\calG_{\sigma}(k)$ is a step function with a transient at $k = 0.5$ as shown in \fref{fig: Ps G0}:
\begin{align*}
\calG_{\sigma}(0)=0, \quad\mbox{and}\quad \calG_{\sigma}(1)=1,
\end{align*}
and $\calG_{\sigma}(k) = 1$ for $k \ge 2$. When $\sigma>0$, $\calG_{\sigma}(k)$ will not be a step function but for small $\sigma$, $\calG_{\sigma}(k)$ can still be approximated by a linear function. Therefore, for small $\sigma$, it holds that
\begin{align*}
\calG_{\sigma}(0)=\delta, \quad\mbox{and}\quad \calG_{\sigma}(1)=1-\delta,
\end{align*}
and $\calG_{\sigma}(k)=1$ for $k \ge 2$. The value of $\delta$ is
\begin{equation}
\delta = \calG_{\sigma}(0) = \Phi\left(\frac{0-q}{\sigma}\right) = \Phi\left(\frac{-0.5}{\sigma}\right).
\end{equation}
For example, if $\sigma = 0.2$, then $\delta \approx 0.0062$.

\begin{figure}[h]
\centering
\begin{tabular}{cc}
\hspace{-1ex}\includegraphics[width=0.475\linewidth]{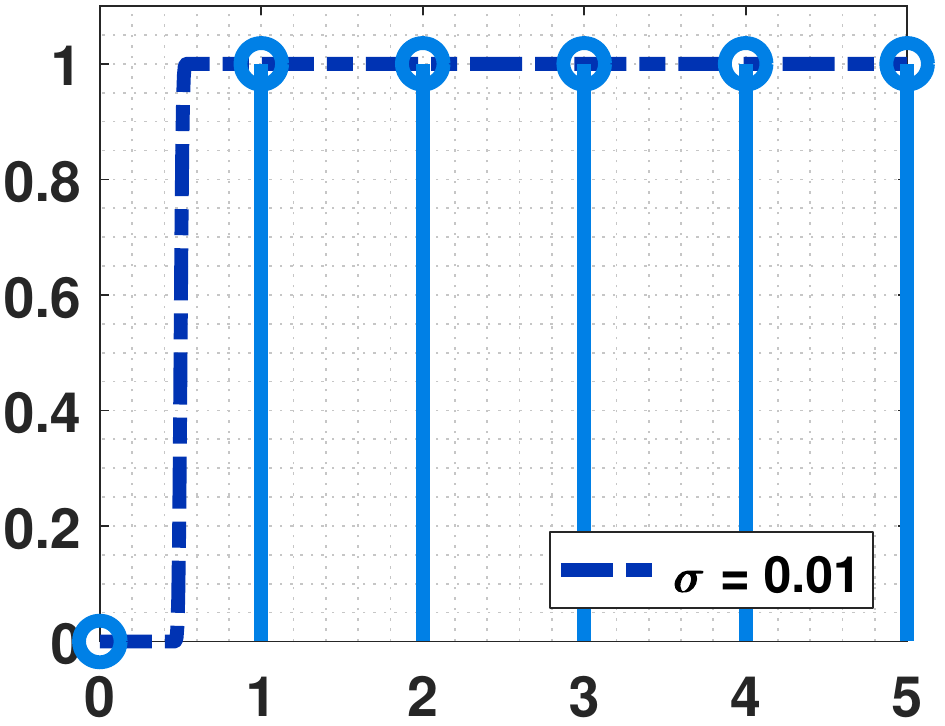}&
\hspace{-2ex}\includegraphics[width=0.475\linewidth]{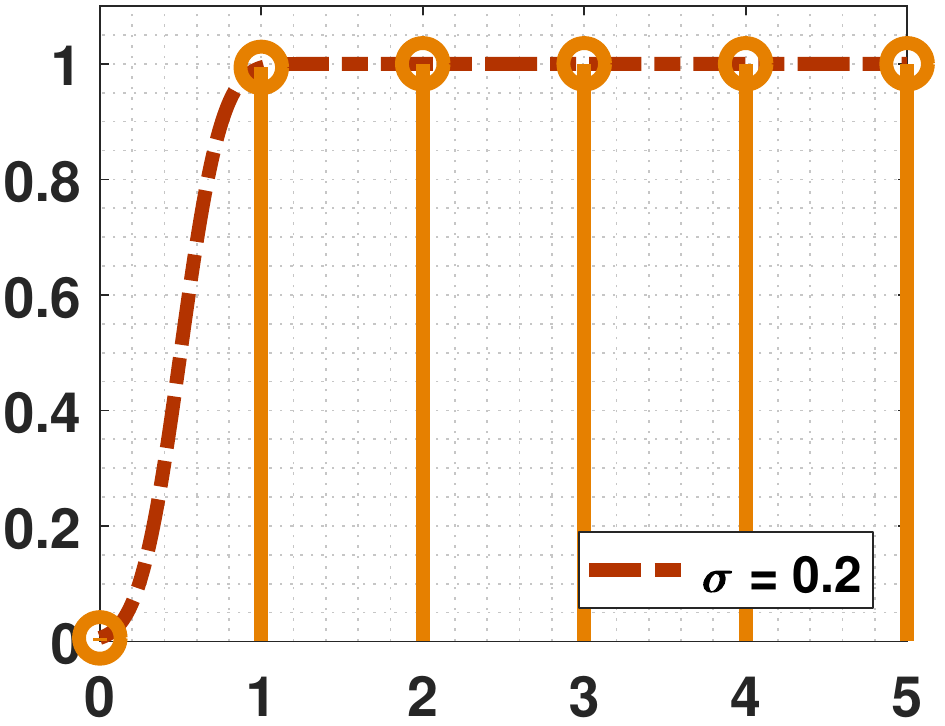}
\end{tabular}
\caption{\textbf{Pictorial illustration of $\calG_\sigma(k)$ in \eref{eq: p_Y(1) for optimal N}}. The pair of subfigures here shows $\calG_\sigma(k)$ for $\sigma = 0.01$ and $\sigma = 0.2$, respectively, assuming $\theta/N = 1$.}
\label{fig: Ps G0}
\end{figure}

Suppose that $\sigma$ is reasonably small so that $\delta$ is small. Then, the probability of getting a one is
\begin{align*}
p_Y(1)
&= \calP_{\theta}(0)\calG_\sigma(0) + \calP_\theta(1)\calG_\sigma(1) + \sum_{k=2}^{\infty}\calP_\theta(k)\calG_\sigma(k)\\
&= e^{-\frac{\theta}{N}} \delta + \frac{\theta}{N}e^{-\frac{\theta}{N}}(1-\delta) + \sum_{k=2}^{\infty}\calP_\theta(k)\calG_\sigma(k)\\
&= \delta e^{-\frac{\theta}{N}} \left(1-\frac{\theta}{N}\right) + \underset{=1-e^{-\frac{\theta}{N}}}{\underbrace{\sum_{k=1}^{\infty}\calP_\theta(k)\calG_\sigma(k)}}\\
&= 1-e^{-\frac{\theta}{N}} \left(1 + \frac{\theta \delta}{N} - \delta\right).
\end{align*}
Typically, the per-frame exposure $\theta/N$ is small. When multiplied with a small constant $\delta$, the term $\frac{\theta \delta}{N}$ can be dropped. This leaves an expression for $p_Y(1)$:
\begin{equation}
p_Y(0) = \omega e^{-\frac{\theta}{N}}, \quad p_Y(1) = 1-\omega e^{-\frac{\theta}{N}},
\label{eq: omega}
\end{equation}
where $\omega = 1-\delta = 1-\Phi(-0.5/\sigma)$. For example, if $\sigma = 0.2$, then $\delta \approx 0.0062$ and $\omega \approx 0.9938$.

Substituting these into \eref{eq: SNR_N}, one can show that
\begin{align*}
\mu_N = p_Y(1) = 1-\omega e^{-\frac{\theta}{N}}, \quad\mbox{and}\quad \frac{d\mu_N}{d\theta} = \frac{\omega}{N} e^{-\frac{\theta}{N}}.
\end{align*}
Hence, the SNR is
\begin{align}
\text{SNR}_{\text{QIS},N}(\theta)
&= \frac{\sqrt{N}\theta}{\sqrt{(1-\omega e^{-\frac{\theta}{N}})(\omega e^{-\frac{\theta}{N}})}} \cdot \frac{\omega}{N} e^{-\frac{\theta}{N}} \notag \\
&= \frac{\theta}{\sqrt{N}}\cdot \sqrt{\frac{1}{\frac{1}{\omega}e^{\frac{\theta}{N}}-1 }}.
\end{align}
This proves the lemma.
\end{proof}

The following is a new theoretical result that summarizes the optimal $\theta$ and $N$ which maximizes this SNR.
\begin{theorem}
\label{thm: theta^* noisy}
Consider the approximated SNR for the noisy case:
\begin{align}
&\log \text{SNR}_{\text{QIS},N}(\theta) \notag \\
&\qquad \qquad = -\frac{1}{2}\log N + \log \theta - \frac{1}{2}\log\left(\frac{1}{\omega} e^{\frac{\theta}{N}}-1\right).
\label{eq: log SNR omega}
\end{align}
The optimal $\theta$ that maximizes $\log \text{SNR}_{\text{QIS},N}$ is
\begin{equation}
\theta^* = \left[W(-2\omega e^{-2}) + 2\right] N,
\end{equation}
where $\omega = 1-\Phi\left(\frac{-0.5}{\sigma}\right)$ is a fixed constant independent of $\theta$ and $N$. For a fixed $\theta$, the optimal $N$ that maximizes the SNR is
\begin{equation}
N^* = \frac{\theta}{1+W(-\omega e^{-1})}.
\end{equation}
\end{theorem}

\begin{proof}
Taking the derivative of the log SNR defined in \eref{eq: log SNR omega}, with respect to $\theta$, will give
\begin{equation*}
\frac{d}{d\theta}\log \text{SNR} = \frac{1}{\theta} - \frac{\frac{1}{\omega N}e^{\frac{\theta}{N}}  }{2\left(\frac{1}{\omega}e^{\frac{\theta}{N}}-1\right)}.
\end{equation*}
Setting this equation to zero will lead to $(\frac{\theta}{N}-2)e^{\frac{\theta}{N}-2} = -2\omega e^{-2}$. Rearranging the terms yields $\theta^* = \left[W(-2\omega e^{-2}) + 2\right] N$.

For the second part of the theorem, the derivative of $\log \text{SNR}$ with respect to $N$ is
\begin{align*}
\frac{d}{dN}\log \text{SNR} = \frac{-1}{2N} + \frac{\frac{1}{N^2\omega}e^{\frac{\theta}{N}\theta}}{2(\frac{1}{\omega}e^{\frac{\theta}{N}}-1)}.
\end{align*}
Rearranging the terms will yield $(\frac{\theta}{N}-1) e^{\frac{\theta}{N}-1} = -\omega e^{-1}$. Hence, the optimal $N$ is
\begin{equation*}
N^* = \frac{\theta}{1+W(-\omega e^{-1})},
\end{equation*}
and this completes the proof.
\end{proof}

The validity of the theoretical derivations can be verified through two observations:
\begin{itemize}
\item As a special case where $\sigma = 0$, then $\omega = 1$. In this case $W(-\omega e^{-1}) = -1$ and so $N^* = \infty$.
\item For any fixed $\sigma$, the constant $\omega = 1-\Phi(-0.5/\sigma)$ is fixed. Then $N^* = \frac{\theta}{1+W(-\omega e^{-1})}$ means that $N^*$ scales \emph{linearly} with $\theta$. The scaling constant for several commonly used read noise levels is shown in Table~\ref{table: N}.
\end{itemize}

\begin{table}[h]
\centering
\caption{The proportional constant $1/(1+W(-\omega e^{-1}))$ as a function of the read noise $\sigma$.}
\begin{tabular}{ccccccc}
\hline
\hline
$\sigma$ & 0 & 0.1 & 0.15 & 0.2 & 0.25 & 0.3 \\
\hline
$\frac{1}{1+W(-\omega e^{-1})}$    & $\infty$ & 1321 &  34.5 & 9.30 & 5.01 & 3.56\\
\hline
\end{tabular}
\vspace{-2ex}
\label{table: N}
\end{table}

\begin{figure}[h]
\centering
\includegraphics[width=\linewidth]{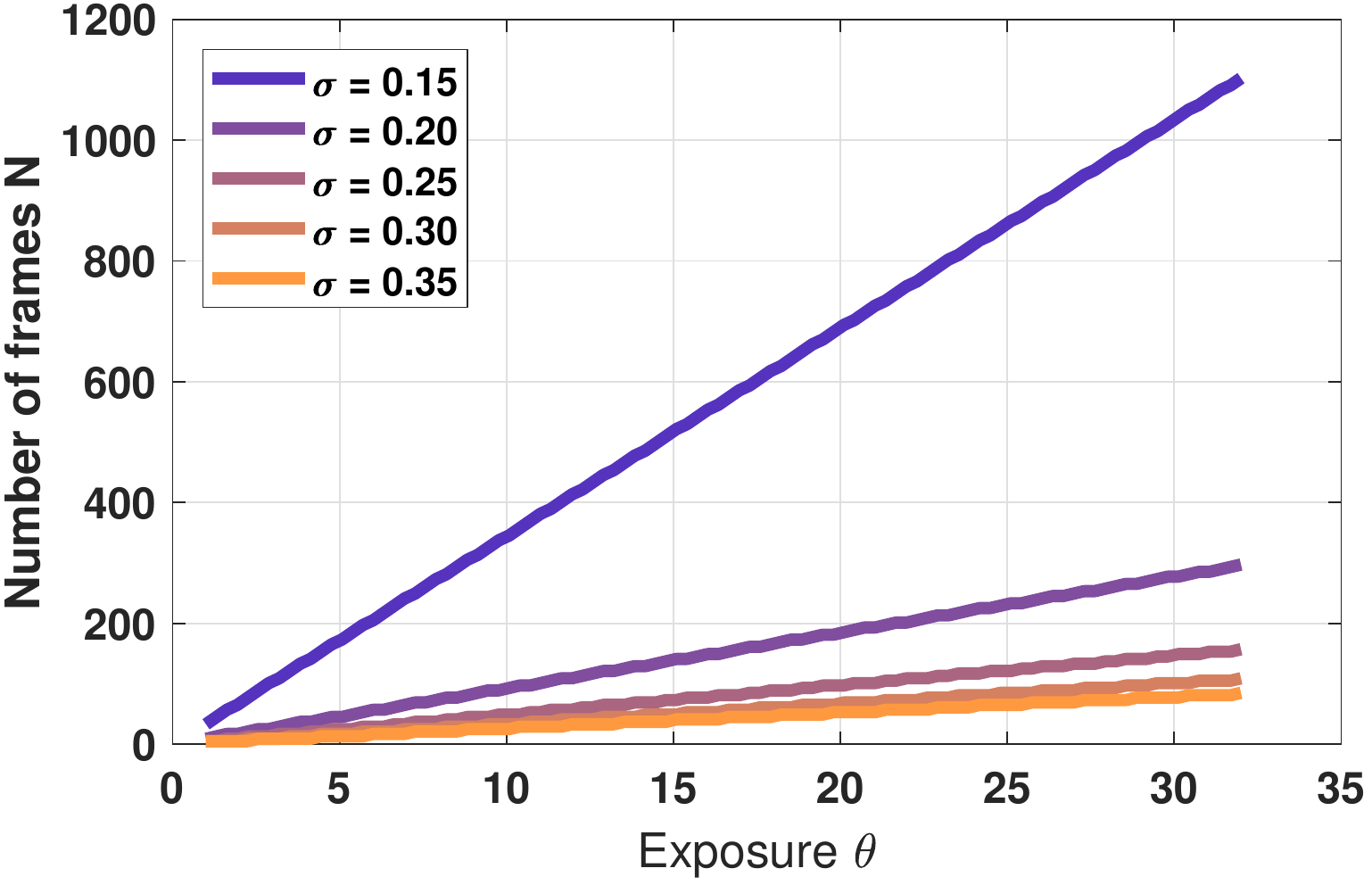}
\vspace{-4ex}
\caption{\textbf{Optimal number of frames $N$} For a fixed $\theta$, the optimal number of frames $N$ scales linearly with the quanta exposure. The slope can be theoretically determined by $1/(1+W(-\omega e^{-1}))$.}
\vspace{-1ex}
\label{fig: Number of Frames 2}
\end{figure}

\fref{fig: Number of Frames 2} shows the optimal number of frames $N^*$ as a function of the exposure $\theta$, evaluated at different read noise $\sigma$. For example, when the read noise is $\sigma = 0.2$, the optimal number of frames $N^*$ is $N^* \approx 9.30 \times \theta$. So, if $\theta = 20$, then $N^* = 186$.

To summarize this section, two key analytic expressions $\theta^*$ and $N^*$ that are missing the literature are now derived. The theoretically predicted $N^*$ match with the numerical simulations. If the read noise is $\sigma = 0$, the optimal $N^*$ is infinity. If the read noise is not zero, the optimal $N^*$ is a finite number. Operating at $N^*$ will maximize the SNR.

\section{Benefit 3: Dynamic Range}
As shown in \fref{fig: low-light}, a one-bit sensor (at $N = 4096$) offers a wider dynamic range compared to a CIS. Where does this come from? The questions to be answered in this section are the following: (1) Is it possible to analytically write down an equation for the dynamic range given the frame rate $N$ and read noise $\sigma$? (2) If exposure bracketing is used, is it possible to theoretically write down an equation of the improvement in the dynamic range?

\subsection{Definition and Mathematical Tools}
\begin{figure}[h]
\centering
\includegraphics[width=\linewidth]{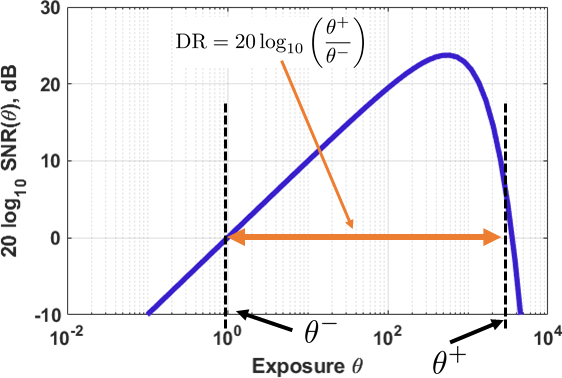}
\vspace{-4ex}
\caption{\textbf{Definition of the dynamic range}. The dynamic range (DR) is defined as the ratio between $\theta^+$ and $\theta^-$, expressed in the log-scale.}
\label{fig: dynamic range}
\end{figure}

The dynamic range of an image sensor is defined as the width of the interval of the exposure (measured in log-scale) within which the SNR is above 1, as shown in \fref{fig: dynamic range}.

\begin{definition}[Dynamic Range \cite{Fossum_Modeling_2013}]
Let $\theta^-$ and $\theta^+$ be two points on the SNR such that
\begin{equation}
\text{SNR}(\theta^+) = \text{SNR}(\theta^-) = 1 \;\; \text{(or 0 dB)},
\end{equation}
where $\theta^- \le \theta^+$. Then, the dynamic range of the sensor is defined as the width
\begin{align}
\text{DR}
&= 20\log_{10}\left(\frac{\theta^+}{\theta^-}\right)
= \underset{\text{width of interval in log space}}{\underbrace{20\log_{10}\theta^+ -  20\log_{10}\theta^-}}.
\end{align}
\end{definition}

To analytically derive an expression of the dynamic range, a variant of the Lambert-W function is needed. Note that this is a customized modification for this specific problem, which does not seem to be mentioned in the literature.
\begin{definition}[Modified Lambert-W function]
Let $0 < c \le 1$ and $0 < \omega \le 1$ be two parameters. The modified Lambert-W functions $V_\omega^+(c)$ and $V_\omega^-(c)$ are the solutions to the equation:
\begin{equation}
\omega\left[\left(\frac{x}{c}\right)^2 + 1\right]e^{-x} = 1,
\label{eq: lambert modified}
\end{equation}
where $V_\omega^-(c) \le V_\omega^+(c)$.
\end{definition}

\begin{figure}[h]
\centering
\includegraphics[width=\linewidth]{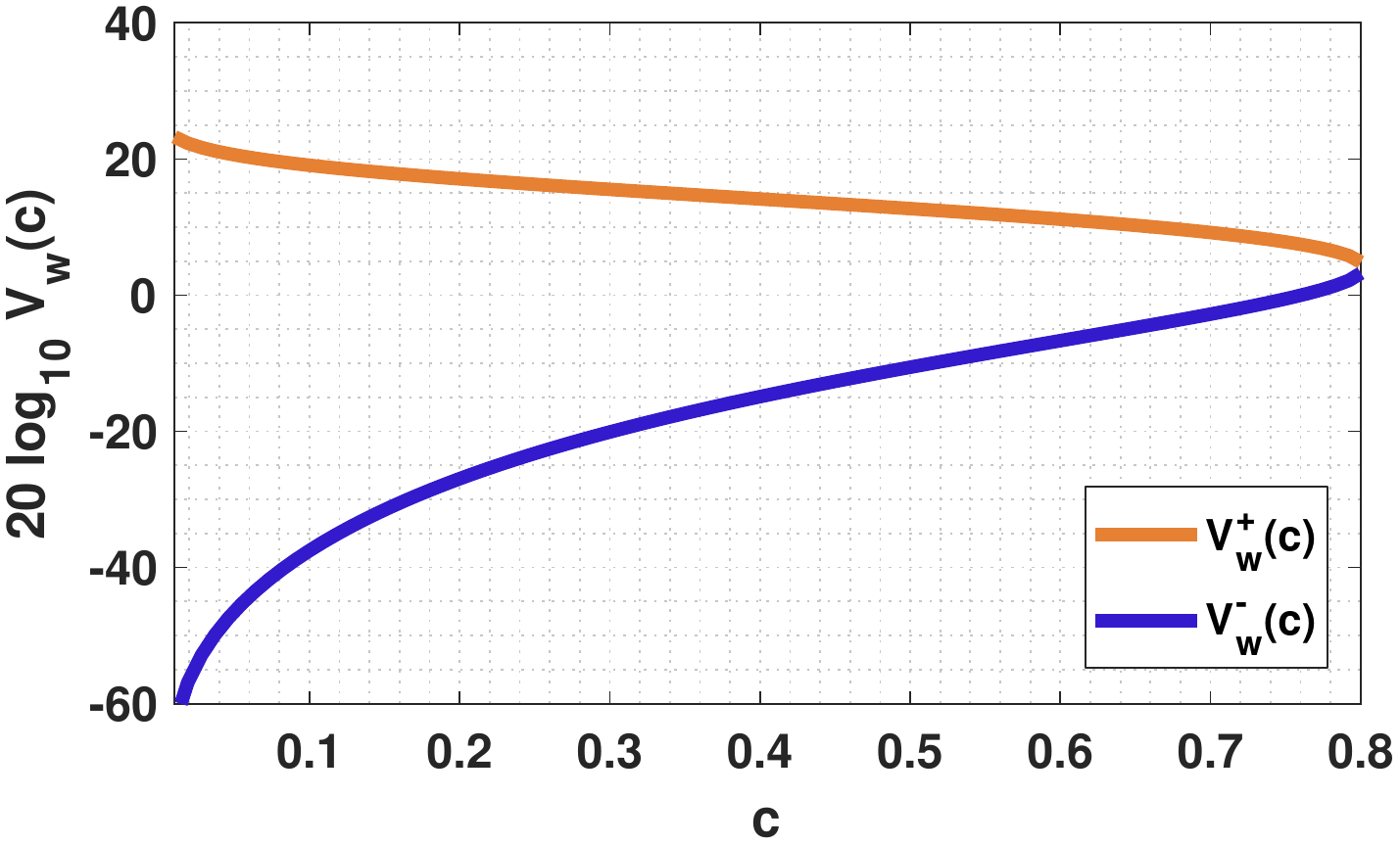}
\vspace{-4ex}
\caption{\textbf{Modified Lambert-W functions}. This figure shows the two modified Lambert-W functions $V_\omega^+(c)$ and $V_\omega^-(c)$ for $\sigma = 0.19$ (so that $\omega = 0.9958$).}
\label{fig: V(c)}
\end{figure}

Like the original Lambert-W function, the modified Lambert-W function cannot be written in terms of elementary functions. To evaluate the modified Lambert-W function, one can first numerically solve \eref{eq: lambert modified} using standard root-finding packages (e.g., \texttt{fzero} in MATLAB) and construct a look-up table for each $\omega$. \fref{fig: V(c)} shows the two modified Lambert-W functions $V_\omega^-(c)$ and $V_\omega^+(c)$ for the case $\sigma = 0.19$ so that $\omega = 0.9958$. For convenience (because the dynamic range is expressed in $20\log_{10}$, the functions $V_\omega^-(c)$ and $V_\omega^+(c)$ are also plotted in the $20\log_{10}$ scale.

\subsection{Analytic Equation of the Dynamic Range}
Analysis of the dynamic range in the past has been limited to empirical studies, i.e., plot the SNR curve and then find the width using a numerical method. The theorem below provides an equation to the dynamic range, expressed in terms of the modified Lambert-W function.

\begin{theorem}
\label{thm: dynamic range}
Consider a one-bit image sensor with a read noise of $\sigma$ and exposure $\theta$. Define $\omega = 1-\Phi(-0.5/\sigma)$ according to \eref{eq: omega}, which is a parameter representing the influence of the read noise to the SNR. For any fixed $N$, the dynamic range of $\text{SNR}_{\text{QIS},N}$ is
\begin{equation}
\text{DR}_{\text{QIS},N} = 20\log_{10} \left( \frac{V_\omega^+\left(\frac{1}{\sqrt{N}}\right)}{V_\omega^-\left(\frac{1}{\sqrt{N}}\right)} \right),
\label{eq dynamic range expression}
\end{equation}
where $V_\omega^+(c)$ and $V_\omega^-(c)$ are the modified Lambert-W function defined in \eref{eq: lambert modified}.
\end{theorem}

\begin{proof}
Recall the definition of $\text{SNR}_{\text{QIS},N}$, and set $\text{SNR}_{\text{QIS},N}(\theta) = 1$:
\begin{align*}
\text{SNR}_{\text{QIS},N}(\theta) = \frac{1}{\sqrt{N}} \cdot \frac{\theta}{\sqrt{\frac{1}{\omega}e^{\frac{\theta}{N}} - 1}} = 1.
\end{align*}
Rearranging the term will yield
\begin{equation*}
\frac{\frac{\theta}{N}}{\sqrt{\frac{1}{\omega}e^{\frac{\theta}{N}} - 1}} = \frac{1}{\sqrt{N}}.
\end{equation*}
Let $x = \frac{\theta}{N}$, and $c = \frac{1}{\sqrt{N}}$, it follows that the equation is identical to
\begin{align*}
\frac{x}{\sqrt{\frac{1}{\omega}e^x -1}} = c,
\end{align*}
which take the same form as the modified Lambert-W function. Therefore, the solution is
\begin{align*}
\frac{\theta^{\pm}}{N} = V_\omega^{\pm}\left(\frac{1}{\sqrt{N}}\right),
\end{align*}
where the superscript $\pm$ means that there is a pair of solutions. Substituting the result into the definition of the dynamic range yields
\begin{align*}
\text{DR}_{\text{QIS},N}
&= 20\log_{10}\left(\frac{\theta^+}{\theta^-}\right)
= 20\log_{10}\left(\frac{V_\omega^+\left(\frac{1}{\sqrt{N}}\right)}{V_\omega^-\left(\frac{1}{\sqrt{N}}\right)}\right),
\end{align*}
which is identical to \eref{eq dynamic range expression}.
\end{proof}

\begin{figure}[h]
\centering
\includegraphics[width=\linewidth]{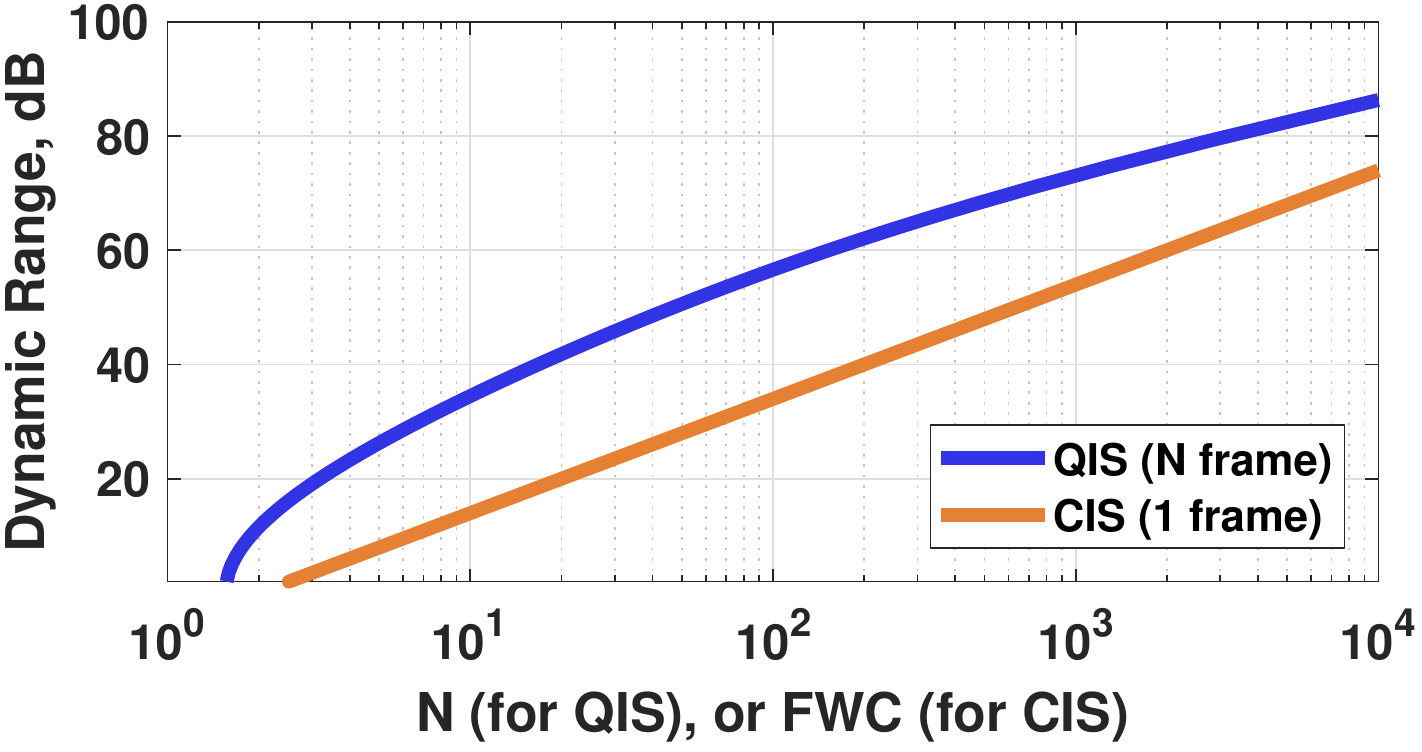}
\vspace{-4ex}
\caption{\textbf{Dynamic range as a function of $N$}. The QIS curve is computed using \eref{eq dynamic range expression}, assuming that $\sigma = 0.19$. For the CIS curve, the full well capacity (FWC) is assumed to be $N$, with a read noise of $\sigma = 2$. }
\label{fig: dynamic range vs N}
\end{figure}

\fref{fig: dynamic range vs N} depicts a pictorial illustration of Theorem~\ref{thm: dynamic range}, for the case where $\sigma = 0.19$ (so that $\omega = 0.9958$.) Three observations are worth of noting:
\begin{itemize}
\item The dynamic range is completely determined by the number of frames $N$ and the read noise $\sigma$, and is independent of $\theta$. The reason is that the dynamic range is defined by the roots $\theta^+$ and $\theta^-$, which are functions of $N$ and $\sigma$ but not $\theta$.
\item The dynamic range grows as $N$ increases. Intuitively, for a small read noise such as $\sigma = 0.2$, Theorem~\ref{thm: theta^* noisy} implies that the optimal $N$ is quite large. For such a large $N$, the per-frame exposure will be short, and so the sensor will be difficult to saturate. This pushed $\theta^+$ to a high value. For $\theta^-$, the low read noise $\sigma$ implies that the per-frame SNR is acceptable. As such, $\theta^-$ can be pushed to a low value. The combination of the two makes the dynamic range increase.
\item The result predicted by Theorem~\ref{thm: dynamic range} is based on the sensor alone using $N$ consecutive frames. There is no carefully designed exposure bracketing to extend the dynamic range. If exposure bracketing is included, the range will be even wider.
\end{itemize}

\subsection{QIS Offers More Dynamic Range than CIS}
How does the dynamic range of the one-bit QIS compare to that of a CIS? For a fair comparison, the analysis is based on Section 3 where the one-bit sensor uses $N$ binary frames while the CIS uses one frame. The full well capacity (FWC) of the CIS is adjusted so that $\text{FWC} = N$ --- If a QIS uses $N = 4000$ binary frames, then equivalently, the CIS should be able to store up to 4000 electrons and so the FWC is 4000.

The following result is known to the sensor's community but a formal proof cannot be found. The proof below is given for completeness.
\begin{theorem}
Consider a CIS with a full well capacity FWC and a read noise $\sigma$. Assume that within a fixed integration time the CIS take one frame. The dynamic range is
\begin{equation}
\text{DR}_{\text{CIS},1} = 20\log_{10}\left(\frac{\text{FWC}}{\sigma}\right).
\end{equation}
\end{theorem}

\begin{proof}
Recall the SNR of the CIS:
\begin{align*}
\text{SNR}_{\text{CIS},1}
=
\begin{cases}
\frac{\theta}{\sqrt{\theta + \sigma^2}}, &\;\; \theta \le \text{FWC},\\
0, &\;\; \theta > \text{FWC}.
\end{cases}
\end{align*}
Because of the sharp cutoff, the upper limit of the dynamic range is FWC, i.e., $\theta^+ = \text{FWC}$. So it remains to determine the lower limit $\theta^-$. To evaluate the lower limit, let $\text{SNR} = 1$ (or 0dB). This will give $\frac{\theta}{\sqrt{\theta + \sigma^2}} = 1$, which can then be written as
\begin{align*}
\theta^2 - \theta - \sigma^2 = 0.
\end{align*}
The solution to this quadratic equation is
\begin{align*}
\theta^- = \frac{1}{2} + \frac{1}{2}\sqrt{1+4\sigma^2} \approx \sigma.
\end{align*}
The approximation is due to the first-order Taylor expansion. Since the typical read noise of CIS is $\sigma = 2$ or more, the approximation is usually acceptable. Note also that the quadratic equation has two roots. The positive root is picked because the exposure cannot be negative.

Putting these together, the dynamic range is
\begin{align*}
\text{DR}_{\text{CIS},1} = 20\log_{10}\left(\frac{\theta^+}{\theta^-}\right) = 20\log_{10} \left(\frac{\text{FWC}}{\sigma}\right),
\end{align*}
which completes the proof.
\end{proof}

Referring back to \fref{fig: dynamic range vs N}, the dynamic range of a CIS can be compared to a one-bit QIS. In this particular configuration where QIS uses $\sigma = 0.19$ and CIS uses $\sigma = 2$, the dynamic range changes as a function of $N$ (or equivalently FWC). For very small or very large $N$, the gap between CIS and a one-bit QIS is small. However, for an $N$ that is in the middle, the gap in the dynamic range can be as large as 20dB.

\subsection{Additional DR Offered by Exposure Bracketing}
A very important characteristic of the one-bit QIS is the exposure bracketing capability. Modern CIS can also do exposure bracketing by dividing the overall exposure into brackets of short and long exposures. However, because of the read noise and quantization of CIS, the shortest bracket cannot be too short. The rapid sampling rate of the one-bit QIS allows it to do exposure bracketing on the smallest interval that a CIS can do.

\begin{figure}[h]
\centering
\begin{tabular}{c}
\includegraphics[width=\linewidth]{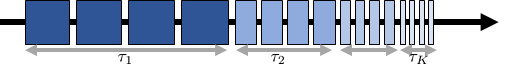}\\
(a) One-bit QIS + bracketing: $K$ brackets, $N$ frames each \\
\includegraphics[width=\linewidth]{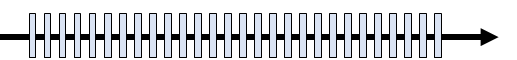}\\
(b) One-bit QIS: $NK$ frames\\
\includegraphics[width=\linewidth]{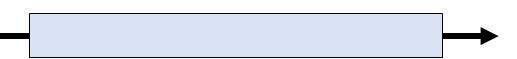}\\
(c) CIS: One frame
\end{tabular}
\vspace{-2ex}
\caption{\textbf{Three exposure-bracketing schemes}: (a) A one-bit QIS that uses $K$ brackets and each bracket uses $N$ binary frames. (b) A one-bit QIS that uses $KN$ short exposure frames. (c) A CIS that uses one frame. All three configurations have the same total duration.}
\label{fig: three brackets}
\end{figure}

To analyze the influence of the exposure bracketing of one-bit QIS, it is necessary to clarify the three comparisons as shown in \fref{fig: three brackets}. In \fref{fig: three brackets}(a), the one-bit QIS uses $K$ brackets to capture a scene where each bracket uses $N$ binary frames. The alternative is \fref{fig: three brackets}(b), where the same one-bit QIS uses $KN$ binary frames during the same total integration time. In \fref{fig: three brackets}(c), the sensor is now a CIS that can only use one frame of a high bit-depth. The question is which configuration will produce the largest dynamic range.

Consider the configuration in \fref{fig: three brackets}(a). Suppose that the total duration is $T$. Within this $T$, there are $K$ brackets $\{\tau_1,\tau_2,\ldots,\tau_K\}$. Without loss of generality, assume that $\tau_k$'s are arranged in the descending order, and define $\tau_{\text{max}} = \tau_1$ and $\tau_{\text{min}} = \tau_K$. The following theorem describes the overall dynamic range of the configuration.

\begin{theorem}
Let $\tau_1,\ldots,\tau_K$ be $K$ exposure brackets of a one-bit QIS. Each bracket produces $N$ binary frames. Then, the overall dynamic range is
\begin{equation}
\text{DR}_{\text{bracket}}
= \underset{\text{one-bit sensor}}{\underbrace{20\log_{10} \left( \frac{V_\omega^+(1/\sqrt{N})}{V_\omega^-(1/\sqrt{N})} \right) }} + \underset{\text{exposure bracketing}}{\underbrace{20\log_{10} \left(\frac{\tau_{\max}}{\tau_{\min}}\right)}},
\end{equation}
where $V_\omega^+(c)$ and $V_\omega^-(c)$ are defined in \eref{eq: lambert modified}.
\label{thm: bracketing}
\end{theorem}

\begin{proof}
Consider the SNR for the $k$th bracket. During the $k$th bracket, the integration time is $\tau_k$. The integration time will amplify/attenuate the exposure seen by the sensor from $\theta$ to $\tau_k \theta$. Therefore, the SNR becomes
\begin{equation}
\text{SNR}_{\text{QIS}, \tau_k}(\theta) = \frac{1}{\sqrt{N}}\cdot \frac{\tau_k \theta}{\sqrt{\frac{1}{\omega} e^{\frac{\tau_k \theta}{N}}-1}}.
\end{equation}
Equating $\text{SNR}_{\text{QIS}, \tau_k}(\theta) = 1$ will give two solutions:
\begin{align*}
\theta^-_{\tau_k} = \frac{N}{\tau_{k}}V_\omega^-\left(\frac{1}{\sqrt{N}}\right), \quad\mbox{and}\quad
\theta^+_{\tau_k} = \frac{N}{\tau_{k}}V_\omega^-\left(\frac{1}{\sqrt{N}}\right),
\end{align*}
which defines the dynamic range for the $k$th bracket.

Since there are $K$ brackets, the merged bracket should use the smallest of $\theta^-_{\tau_k}$ (so that it can capture the darkest scene) and the largest of $\theta^+_{\tau_k}$ (for the brightest scene). Therefore, by defining
\begin{align*}
\theta^- &= \min\{\theta_{\tau_k}^- \;|\; k = 1,\ldots,K\} = \frac{N}{\tau_{\max}}V_\omega^-\left(\frac{1}{\sqrt{N}}\right),\\
\theta^+ &= \max\{\theta_{\tau_k}^+ \;|\; k = 1,\ldots,K\} = \frac{N}{\tau_{\min}}V_\omega^-\left(\frac{1}{\sqrt{N}}\right),
\end{align*}
where $\tau_{\text{min}} = \tau_K$ and $\tau_{\text{max}} = \tau_1$, the overall dynamic range is
\begin{align*}
\text{DR}_{\text{bracket}}
&= 20\log_{10} \theta^+ - 20\log_{10} \theta^-\\
&= 20\log_{10} \left( \frac{V_\omega^+(1/\sqrt{N})}{V_\omega^-(1/\sqrt{N})} \right)  +  20\log_{10} \left(\frac{\tau_{\max}}{\tau_{\min}}\right).
\end{align*}
This completes the proof.
\end{proof}

The analytic expression of the dynamic range derived in Theorem~\ref{thm: bracketing} is important in several ways. First, it says that the dynamic ranges offered by the exposure bracketing and that of the one-bit sensor alone are \emph{decoupled}. This explicit relation is a new finding. Second, the decoupling implies that the bigger the ratio $\tau_{\text{max}}/\tau_{\text{min}}$ is, the wider the \emph{additional} dynamic range one can obtain.

\begin{figure}[h]
\centering
\includegraphics[width=\linewidth]{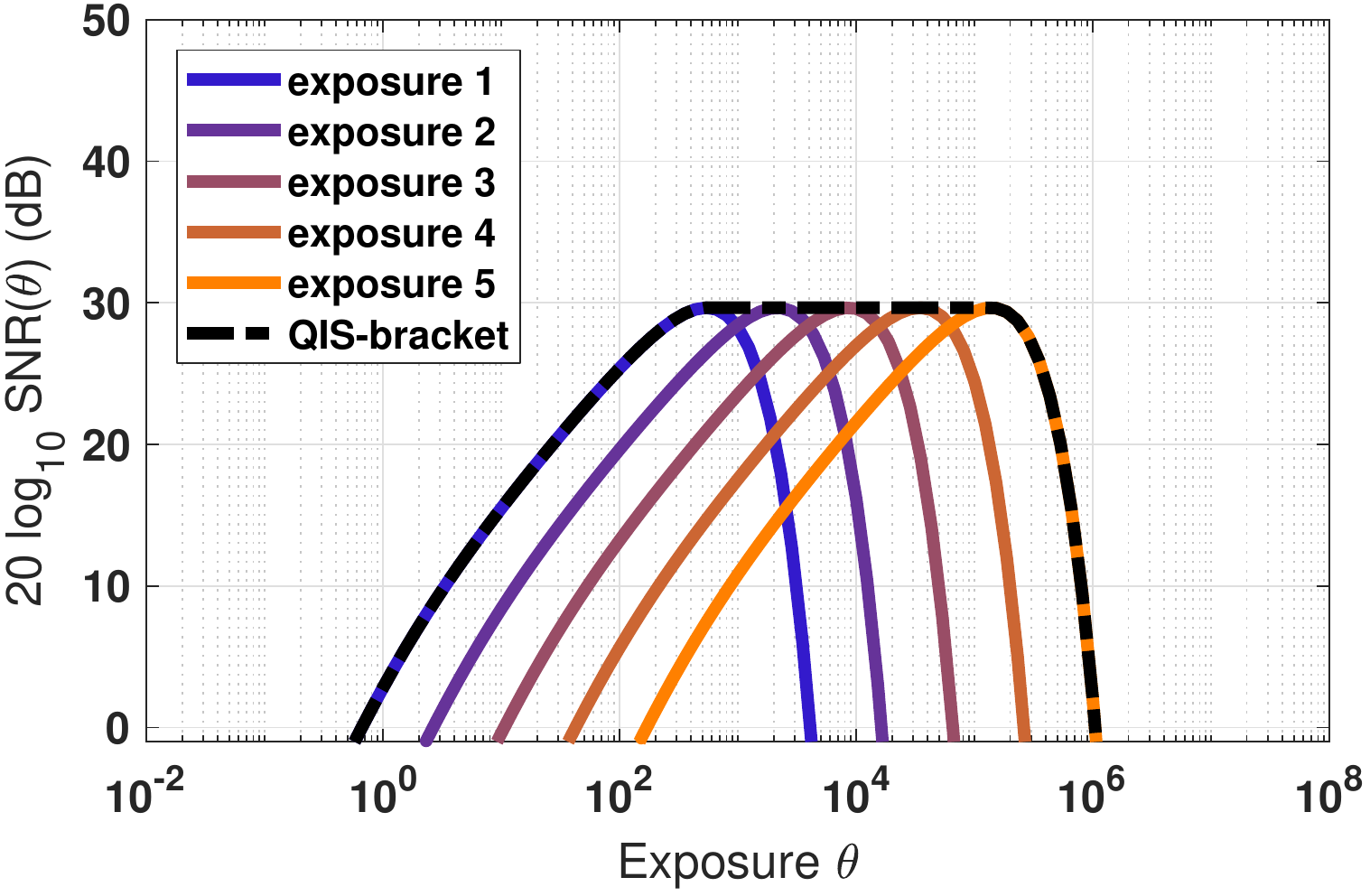}
\vspace{-4ex}
\caption{\textbf{Overall SNR by exposure bracketing}. The dynamic range predicted by Theorem~\ref{thm: bracketing} suggests a decoupling between the one-bit sensor and the bracketing integration time. Each bracket has the same dynamic range, but $\tau_k$ changes the horizontal displacement of the SNRs.}
\label{fig: bracketing SNR}
\end{figure}

\fref{fig: bracketing SNR} shows an example where there are $K = 5$ brackets: $\{4,1,\tfrac{1}{4}, \tfrac{1}{16}, \tfrac{1}{64}\}$. Each bracket produces $N = 1000$ binary frames. As can be seen in the figure, the dynamic range of each bracket remains a constant no matter which $\tau_k$ is used. They are differed by a horizontal translation. The expanded dynamic range is determined by how much translation the individual exposures have experience from $\tau_{\text{max}}$ to $\tau_{\text{min}}$.

For this particular example, the read noise is $\sigma = 0.19$ (so that $\omega = 0.9958$. Let $N = 1000$, $\tau_{\text{max}} = 4$ and $\tau_{\text{min}} = 1/64$. Then,
\begin{align*}
20\log_{10} \left( \frac{V_\omega^+(1/\sqrt{N})}{V_\omega^-(1/\sqrt{N})} \right) &= 73.07 \, \text{dB},\\
20\log_{10} \left(\frac{\tau_{\max}}{\tau_{\min}}\right)                         &= 48.16 \, \text{dB},
\end{align*}
and so the overall dynamic range is $\text{DR}_{\text{bracket}}  = 121.24 \, \text{dB}$.

Now consider the configuration in \fref{fig: three brackets}(b) where the same one-bit QIS uses a total of $KN$ binary frames to generate one image. Since there is no bracketing involved, the dynamic range is
\begin{equation}
\text{DR}_{\text{QIS}, KN} = 20\log_{10}\left( \frac{V_\omega^+(1/\sqrt{KN})}{V_\omega^-(1/\sqrt{KN})} \right).
\end{equation}
Substituting $K = 5$ and $N = 1000$, the dynamic range offered by this configuration is $\text{DR}_{\text{QIS}, KN} = 82.51 \,\text{dB}$.

Finally, consider a CIS that uses only one frame to capture the scene. The equivalent full well capacity relative to the other two configurations is that $\text{FWC} = KN$. Thus, the dynamic range is
\begin{equation}
\text{DR}_{\text{CIS}, 1} = 20\log_{10}\left( \frac{\text{FWC}}{\sigma}\right) = 20\log_{10}\left( \frac{KN}{\sigma}\right).
\end{equation}
Assuming $\sigma = 2$, and $\text{FWC} = 5 \times 1000 = 5000$, it follows that $\text{DR}_{\text{CIS}, 1} = 67.96 \, \text{dB}$.

\begin{figure}[h]
\centering
\includegraphics[width=\linewidth]{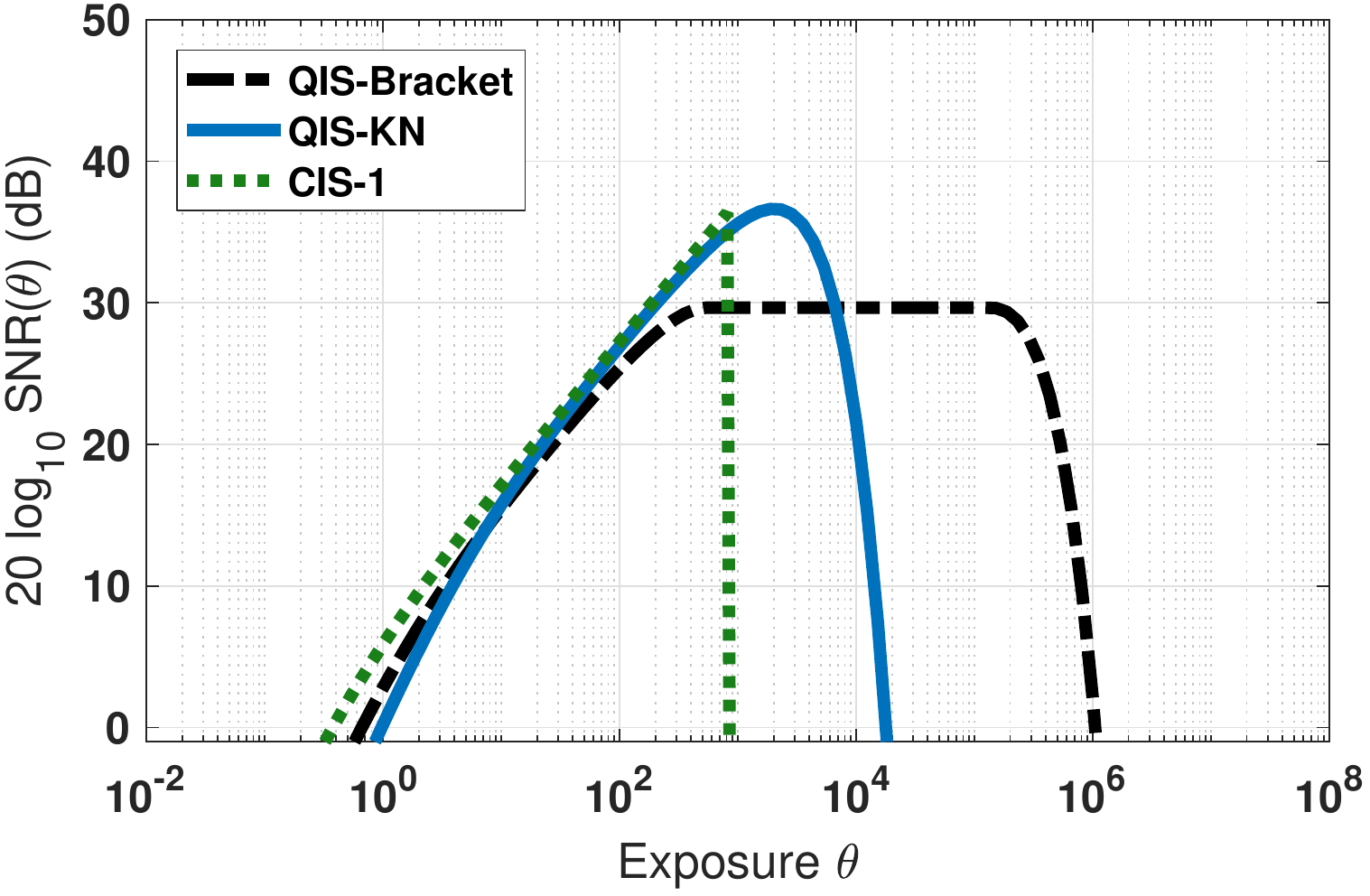}
\vspace{-4ex}
\caption{\textbf{SNR comparisons of the three configurations}. The three curves correspond to the three configurations depicted in \fref{fig: three brackets}. }
\label{fig: bracketing CIS}
\end{figure}

The comparison of the above three configurations is shown in \fref{fig: bracketing CIS}. For configurations (b) \texttt{QIS-KN} and (c) \texttt{CIS-1}, the location of the SNR can be shifted horizontally to the right by using a smaller overall exposure. Using a smaller overall exposure will not change the dynamic range but will help the sensor perform better at high-light (with a compromise at low-light). Without shifting the SNR of \texttt{QIS-KN} and \texttt{CIS-1} to the right, the two SNRs have a better performance at low-light because their overall exposure is larger than $\tau_{\text{max}}$. If they are shifted to the right, then their low-light performance will become much worse than (a) \texttt{QIS-Bracket}.

To summarize, the dynamic range offered by a one-bit QIS with exposure bracketing can offer 121.24 dB on a particular but typical configuration. Compared to a CIS that offers 67.96 dB, the improvement is 53.28 dB. Converting back to the linear scale, this is about $10^{53.28/20} = 461.32$ times improvement of the dynamic range.

\section{Discussions and Conclusion}
The summary of the findings of this paper is as follows:
\begin{itemize}
\item Low-light: The performance gain of the one-bit QIS is due to two facts: (i) it has a lower read noise so that it can differentiate photons; (ii) the binary quantization further reduces the probability of error in the bits. To achieve this performance, the one-bit QIS needs to operate at $\theta \approx 1$ by adjusting the integration time. CIS performs worse because of the combination of read noise and analog-to-digital conversion. However, if the one-bit QIS is not configured properly, CIS can still perform better.
\item Frame rate: The frame rate of the one-bit QIS cannot be arbitrarily high unless the read noise is zero. A high frame rate causes individual frames to suffer from read noise because the per-frame exposure is short. The optimal frame scales linearly with the exposure $\theta$ where the proportional constant is determined by the read noise.
\item Dynamic range: One-bit QIS offers two benefits. First, the nonlinearity of the one-bit statistics gives an improved dynamic range compared to a CIS. Second, exposure bracketing further improves the dynamic range. The two factors are independent to each other.
\end{itemize}

The analysis presented in this paper is complete on its own, but several open questions are worthy of pursuing. First, the Poisson-Gaussian model did not consider the dark current, pixel non-uniformity, defects, and threshold instability. The paper has also not considered the multi-bit scenarios, which would require the analysis of the analog-to-digital conversion and the utilization of the incomplete Gamma function. On the frame rate analysis, an important question is the impact of the scene motion. However, without a tool to decouple the sensor-level analysis and the motion estimation method, the problem would be hard because the conclusion may be restrictive to a particular motion estimation method. Finally, while the benefit of exposure bracketing is proven in this paper, the specific bracketing strategy is yet to be developed, e.g., how many brackets and the duration of each bracket.

\section*{Acknowledgement}
The author thanks Professor Eric Fossum for sharing many thoughts on the manuscript and Purdue Ph.D. students Abhiram Gananasambandam and Yash Sanghvi for discussing the results of the manuscript. The author also thanks the anonymous reviewers for their valuable feedback.

\bibliography{ref}
\bibliographystyle{ieeetr}
\end{document}

%% file: TeX/commands.tex
\ifCLASSINFOpdf
  \usepackage[pdftex]{graphicx}
\else
\fi

\usepackage[colorlinks]{hyperref}
\usepackage{multirow}
\usepackage{times}
\usepackage{amsmath,amssymb}
\usepackage{algorithm}
\usepackage{epstopdf}
\usepackage{subfigure}
\usepackage[dvipsnames]{xcolor}
\usepackage{algorithm}
\usepackage{algorithmic}
\usepackage{cite}
\usepackage{fancyvrb}

\usepackage{bbm}
\newtheorem{theorem}{Theorem}

\newtheorem{lemma}{Lemma}

\newtheorem{definition}{Definition}

\newenvironment{proof}{\paragraph*{Proof}}{\hfill$\square$}

{\begin{list}               %    with flush left bullets.
    {$\bullet$ \hfill}{
        \setlength{\leftmargin}{\parindent}
        \setlength{\parsep}{0.04\baselineskip}
        \setlength{\itemsep}{0.5\parsep}
        \setlength{\labelwidth}{\leftmargin}
        \setlength{\labelsep}{0em}}
    }
{\end{list}}

\providecommand{\eref}[1]{\eqref{#1}}  % call \eqref from amstex
\providecommand{\cref}[1]{Chapter~\ref{#1}}

\providecommand{\fref}[1]{Figure~\ref{#1}}

\providecommand{\E}{\ensuremath{\mathbb{E}}}

\providecommand{\bydef}{\overset{\text{def}}{=}}

\providecommand{\mat}[1]{\ensuremath{\boldsymbol{#1}}}

\providecommand{\calG}{\mathcal{G}}

\providecommand{\calP}{\mathcal{P}}

\providecommand{\mY}{\mat{Y}}

\providecommand{\thetahat}{\widehat{\theta}}

\providecommand{\Var}{\mathrm{Var}}

\newcommand{\argmax}[1]{\mathop{\underset{#1}{\mbox{argmax}}}}